## Title

# Charge-conserving creation and annihilation of topological phase singularities in atomically thin semiconductors

## Authors

Yongjing Wang[1#], Yuhang Wang[1#], Guoshuai Fan[1], Qingqing Luo[1], Yanning Li[1], Lidong Sun[2], Chunguang Hu[1*], & Wanfu Shen[1*]

## Affiliations

1. State Key Laboratory of Precision Measuring Technology and Instruments, School of Precision Instrument and Opto-Electronics Engineering, Tianjin University, Tianjin 300072, China

2. Fundamental Science Department, Hainan Bielefeld University of Applied Sciences, 570100 Hainan, China

## Corresponding author

Correspondence to: cghu@tju.edu.cn; wfshen@tju.edu.cn;

## Abstract

Atomically thin transition metal dichalcogenides provide a promising material platform for optical phase manipulation, as rapid variations in their dielectric functions can give rise to topological phase singularities (PSs). However, PS evolution while preserving topological charge conservation has not yet been investigated in exciton enabled PS systems. Here, using atomically thin $MoS_2$ films on sapphire as a model system, we theoretically and experimentally demonstrate that the evolution of dielectric dispersion determines the creation, trajectory, merging, and annihilation of optical PSs under topological charge conservation. By varying the $MoS_2$ film thickness, we experimentally realize this evolution through the gradual shrinking of the high energy $n-k$ dispersion loop, which modifies its intersection with the zero reflection condition. Our results provide a general strategy for exploring and manipulating topological optical states.

## INTRODUCTION

Nanophotonic phase engineering provides a powerful approach to compact optical systems by tailoring light fields through subwavelength control of optical phase (*1–4*). A particularly powerful route exploits topological phase singularities (PSs) in parameter space where the reflected amplitude vanishes and the optical phase becomes undefined (*5–8*). Around such points, the phase winds by an integer multiple of $2\pi$, endowing the singularity with a non-trivial topological charge that makes it robust against moderate structural imperfections and perturbations (*9–11*). When the parameter trajectory crossing the singularity, the phase undergoes an abrupt $\pi$-phase jump (*12*). This topological protection, together with the diverging phase response near the singularity, underpins both phase engineering and ultrasensitive detection (*13–16*).

In contrast to conventional phase-singular photonic systems based on plasmonic nanostructures (*17*, *18*), metasurfaces (*19*, *20*), photonic crystals (*21*, *22*) and cavity-assisted resonant structures (*23*, *24*), transition metal dichalcogenides (TMDC) monolayers support topological PSs through their intrinsic optical response dominated by exciton resonances without external cavity assistance, extending phase-singular photonics to the single-atomic-layer limit (*25*). Notably, for a linear optical medium, the analyticity of the complex dielectric function $\varepsilon(\omega)$ enforced by the Kramers-Kronig relations, determins the winding sense of the $n - k$ dispersion trajectory around each absorption resonance (*26*, *27*). Consequently, the Lorentzian resonance might induces the dielectric function to trace a sharp loop in the complex permittivity plane that intersects the zero reflection singular line twice, crossing the singular line with opposite winding orientations, which generates a phase singularity pair carrying charges +1 and -1. Conservation of the total topological charge governs singularity creation and annihilation, which typically occur through oppositely charged pairs. However, these processes have not yet been investigated in exciton-enabled PS systems.

In atomically thin TMDCs, most atoms are close to the surrounding interfaces making the optical response sensitive to sample conditions such as the local dielectric environment, defect and synthesis methods (*28–30*). At the same time, the dielectric function itself is complex, with multiple absorption peaks arising from pronounced excitonic resonances as well as higher energy interband transitions associated with enhanced joint density of states near high symmetry critical points (*31*, *32*). For example, in monolayer $MoS_2$ film, the widely observed E peak may involve several nearby critical point transitions rather than a single isolated absorption resonance (*33*). Substrate screening, optically active defects, temperature and synthesis methods can shift or broaden these high energy absorption peaks and redistribute their oscillator strengths (*29*, *34–36*). These variations can modify the absorptive and dispersive components of the dielectric function following the Kramers–Kronig relation (*37*), might reshaping the $n - k$ dispersion trajectory and altering its intersections with the zero reflection singular line, providing a potential route for controlling the evolution of topological PSs.

Here, we demonstrate the creation and annihilation of charge paired topological PSs in atomically thin $MoS_2$ films prepared on sapphire substrates by sulfurizing Mo films, revealing charge conserving evolution of PSs. First, using Kramers–Kronig constrained optical constants and zero reflection analysis, we demonstrate that the formation and evolution of topological PSs are closely related to the intrinsic dielectric dispersion of the material itself. The closed loop formed by the dielectric dispersion in the $(n, k)$ parameter space determines the evolution of PS pairs. Instead of introducing complex optical structures, film thickness is used as a direct parameter to modify the dielectric response of $MoS_2$, thereby reshaping the $n - k$ dispersion loop and realizing the evolution of PSs. By tuning the $MoS_2$ film thickness, an oppositely charged PS pair emerges in the ultraviolet spectral region. With increasing thickness, the two PSs gradually approach each other and eventually annihilate, following the conservation of total topological charge. The phase sensitivity of PSs to refractive index variations was evaluated, with a maximum sensitivity of $9.35 \times 10^4$ deg/RIU obtained at the ultraviolet PS. Our findings provide a route for controlling the evolution and annihilation of topological PSs in atomically thin materials, offering a new perspective for singular optics in two dimensional semiconductors.

## RESULTS

## Theoretical prediction of oppositely charged PS pairs

To reveal how dielectric dispersion controls PS evolution, the zero reflection condition was first established for the air/$MoS_2$/sapphire structure shown in Fig. 1c using the transfer matrix method (*17*, *38*, *39*). The ultrathin $MoS_2$ film is supported on a c-plane sapphire substrate with a roughened backside to suppress backside reflection, allowing the sapphire substrate to be treated as semi infinite in the optical calculation. Following the approach reported previously (*25*), the zero reflection condition was mapped in the reduced $(n, k)$ parameter space, where $n$ and $k$ denote the refractive index and extinction coefficient respectively. At a fixed incidence angle, wavelength and film thickness, the zero reflection condition appears as a continuous singular line, which serves as a criterion for determining whether a material dispersion trajectory can generate PSs (see Supplementary Note 1 for calculation details). The influence of dielectric dispersion on this crossing condition was then examined by modifying the optical constants while preserving their physical correlation following the Kramers–Kronig relation (*37*). Specifically, the extinction coefficient $k$ of monolayer $MoS_2$ was systematically varied to simulate changes in the absorptive component of the dielectric response, and the corresponding refractive index $n$ was recalculated accordingly.

As shown in Fig. 1a, increasing the $k$ perturbation drives the ultraviolet $n - k$ dispersion loop from two crossings with the singular line, through a critical tangent condition, to a state without intersection. These transitions correspond to the creation and annihilation of a PS pair (*24*). This perturbation analysis shows that the additional high energy PS pair emerges only when the ultraviolet $n - k$ dispersion loop enters a regime where it intersects the singular line. When the dispersion loop moves away from this condition, the intersection is lost and the PS pair is annihilated. To further examine whether this connection is reflected in real $MoS_2$ samples, reported $n - k$ dispersions of monolayer $MoS_2$ prepared by different methods, substrate, environments and transfer processes were collected, including as-grown and transferred samples (*36*, *25*, *40*, *41*, *29*, *30*, *42–44*). Fig. S1 compares the $n - k$ dispersion curve of the $MoS_2$ film used in this work with nine reported monolayer $MoS_2$ optical constant datasets. Although these datasets exhibit similar overall dispersion trends, clear variations are observed in dielectric amplitude and high energy optical features. Most reported monolayer $MoS_2$ dispersion curves intersect the zero reflection singular line near the C excitonic region where strong absorption and dielectric dispersion are commonly observed. In contrast, the additional high energy PS pair requires the $n - k$ dispersion loop near the E feature to cross the singular line, many dispersion curves form loops, but only a part of these loops cross the singular line, including the curve obtained in this work. This indicates that the high energy $n - k$ loop is closely related to the dielectric variation associated with the E feature, and that changes in the E feature can affect whether the additional high energy PS pair emerges in atomically thin $MoS_2$ without external cavity assistance.

To clarify the dielectric origin of the high energy $n - k$ loop, the imaginary part of the dielectric function $\varepsilon_2$ of the $MoS_2$ film was fitted with multiple Lorentz oscillators (Fig. 1b; see Supplementary Fig. S6 and Table S1 for fitting details and oscillator parameters). Multiple optical transitions are resolved in the fitted $\varepsilon_2$ spectrum, among which the high-energy E feature may arise from multiple high-energy interband transitions from the highest valence band to the second lowest conduction band near the M/K points(*33*). Therefore, the E feature represents a complex high energy dielectric response that is sensitive to dielectric screening, defects and synthesis conditions, which can indirectly modify its peak position, linewidth and oscillator strength (*45*). These changes reshape the $n - k$ loop near 300 nm, thereby modifying its intersection with the zero-reflection singular line and determining the emergence of the additional PS pair.

Following the simplified $n - k$ analysis, transfer matrix calculations based on the structure in Fig. 1c were extended to the full $(n, k, \theta, \lambda)$ parameter space to further verify the predicted additional PS pair and identify their corresponding spectral and angular positions (see Supplementary Note 1). The wavelength dependent optical dispersion of $MoS_2$ is described by $N(\lambda) = n(\lambda) - ik(\lambda)$. For a given $MoS_2$ thickness, zero reflection occurs when the p-polarized reflection coefficient satisfies $r_p(n, k, \theta, \lambda) = 0$. This requires both the real and imaginary parts of $r_p(n, k, \theta, \lambda)$ to vanish simultaneously, thereby imposing two conditions that determined the zero-reflection wavelength $\lambda(n, k)$ and incidence angle $\theta(n, k)$ for each given combination of $n$ and $k$ (*6*). Plotted in $(n, k, \lambda)$ the parameter space, this condition defines a zero-reflection surface (ZRS), shown by the blue points in Fig. 1d. The $MoS_2$ dispersion curve, shown as the red solid curve, intersects the ZRS to produce three distinct intersection points, marked by yellow stars, confirming the three zero-reflection points predicted from the reduced $n - k$ analysis.

At these points, the reflected amplitude vanishes and the optical phase becomes undefined. Fig. 1e shows the simulated ellipsometric parameter $\Delta$ map in the vicinity of these points revealing the phase winding. In the ultraviolet region, the additional pair of PSs exhibits opposite phase windings of $+2\pi$ and $-2\pi$, corresponding to topological charges of $C = +1$ and $C = -1$, respectively. The PS near the C excitonic region exhibits a $+2\pi$ phase winding, corresponding to $C = +1$. These opposite winding directions originate from the opposite crossing orientations between the closed dispersion loop and the ZRS consistent with the topological properties of closed curves described by the Jordan curve theorem (*12*). Under small perturbations to the dielectric function or film parameters, such PSs can move in parameter space, whereas their annihilation occurs only when opposite charge singularities merge. Therefore, the creation and annihilation of PSs obeys topological charge conservation. The corresponding simulated $\Psi$ map in Fig. 1f further reveal the near zero regions, confirming the locations of the three PSs. These PSs are primarily governed by the intrinsic dielectric dispersion of $MoS_2$ rather than Brewster effect, interference or cavity assisted mechanisms (*46–48*). The $MoS_2$ film is absorptive and only approximately 1 nm thick, so propagation phase accumulation across the film is negligible. Moreover, the sapphire substrate cannot provide the Fabry–Pérot resonances available in $SiO_2$/Si substrates, which amplify the optical response of an atomically thin film in the overall reflection signal (*24*). These considerations suggest that the emergence of the oppositely charged PS pair originates from the dielectric dispersion of atomically thin $MoS_2$.

## Structural and optical characterization of the $MoS_2$ film

Atomically thin $MoS_2$ films were synthesized on sapphire substrates using a two-step method (see Methods and Supplementary Note 2) (*49*). The uniformity of the $MoS_2$ films was confirmed by the combined structural and spectroscopic characterizations shown in Fig.2. The crystal structure of $MoS_2$ is illustrated schematically in Fig. 2a, while Scanning Electron Microscopy (SEM) image reveals a continuous and uniform film over a large area (Fig. 2b). Next, the surface morphology and thickness of the film were characterized by Atomic Force Microscopy (AFM). The AFM image shows a smooth and uniform surface, with a thickness of approximately 1.05 nm (Fig. 2c,d). For comparison, the AFM image of the bare sapphire substrate is also provided in Fig. S3a. Additionally, under 532 nm excitation, the Photoluminescence (PL) spectrum shows a sharp emission peak at 679 nm, corresponding to the direct excitonic transition at 1.826 eV and indicating good optical quality (Fig. 2e). Raman spectroscopy under the same excitation wavelength reveals the characteristic $E_{2g}^{1}$ and $A_{1g}$ vibrational modes of $MoS_2$ at approximately 385.08 and 405.89 $\mathrm{cm}^{-1}$, respectively, with a Raman shift difference of 20.81 $\mathrm{cm}^{-1}$ (Fig. 2f). Furthermore, spatial mapping of the Raman shift difference between the $E_{2g}^{1}$ and $A_{1g}$ modes was performed confirming the thickness uniformity of the $MoS_2$ film (Fig. 2g). The complex optical constants of $MoS_2$ were extracted over a broad spectral range of 210–1000 nm, where the red and blue solid curves denote the refractive index $n$ and extinction coefficient $k$, respectively (Fig. 2h). As expected, excitonic resonances appear at lower energies (*50*), while critical-point-related interband transitions emerge at higher energies, together giving rise to the strong optical response of the $MoS_2$/sapphire system (*31*). These experimentally obtained optical constants were directly used as the material parameters for the subsequent zero-reflection and phase-singularity analysis.

## Measurement of oppositely charged PSs in $MoS_2$ film

To experimentally verify the predicted oppositely charged PS pair in the exciton enabled $MoS_2$/sapphire system, variable angle spectroscopic ellipsometry (SE) was employed (see Methods). SE enables simultaneous measurement of the ellipsometric parameters $\Psi$ and $\Delta$, which are defined by the complex reflection ratio $\rho$: (*51*)

$$\rho = \frac{r_p}{r_s} = \tan\Psi\, e^{i\Delta} \tag{1}$$

Here, $\tan\Psi$ represents the amplitude ratio between the $p$ and $s$ polarized reflection coefficients, and $\Delta$ represents their phase difference. At specific combinations of wavelength and incidence angle, $\Psi$ approaches zero, indicating that $r_p/r_s$ vanishes and the zero reflection condition $r_p = 0$ is satisfied.

Experimentally, the light reflection measured from 1.05 nm thick $MoS_2$ film on a sapphire substrate shows the PSs pair in ultraviolet region. Fig. 3a shows that for incidence angles fixed at 63.6°, 63.9° and 65.5°, $\Psi$ approaches zero at the corresponding wavelengths of 323, 318, and 424 nm consistent with the theoretical predictions. Furthermore, the three $\Psi$ minima are clearly identified in the two-dimensional $\Psi$ map of the wavelength-angle parameter space, obtained by scanning the incidence angle from 62° to 68° in 0.1° steps (Fig. 3b), in good agreement with the spectra in Fig. 3a. However, the $\Psi$ dips alone is insufficient to confirm the presence of the PSs, as the phase response must also be examined. The $\Delta$ map in Fig. 3c reveals phase winding of $+2\pi$ or $-2\pi$ around zero-reflection points, corresponding to nonzero topological charges and confirming the topological nature of the PSs. In the ultraviolet region, the two additional PSs carry opposite charges of $C = +1$ and $C = -1$, whereas the PS near the C excitonic region exhibits a charge of $C = +1$. Close examination of Fig. 3b,c reveals that the PSs near 318 and 323 nm appear less sharply resolved than in the theoretical simulations owing to experimental non-idealities (finite spectral/angular resolution, surface roughness, etc.).

The characteristic phase response of the PSs is further revealed by the extracted $\Psi$ and $\Delta$ spectra at incidence angles near the singularity conditions (Fig. 3d). For the two oppositely charged singularities, spectra were extracted at fixed incidence angles of 63.9° and 63.6°, corresponding to the zero-reflection points near 318 and 323 nm, respectively. For the singularity near 424 nm, spectra were extracted slightly below (65.4°) and above (65.6°) the PS condition at 65.5° ± 0.1° to capture the phase branch transition across the zero-reflection point. In all cases, rapid variations in $\Delta$ are observed in the vicinity of the $\Psi$ dips. By extracting $\Delta$ as a function of incidence angle at fixed wavelengths. As shown in Fig. 3e, $\Delta$ exhibits an abrupt $\pi$ jump upon crossing the singularity condition, consistent with the Heaviside-like phase response of zero-reflection points and providing a clear experimental signature of the PSs (*25*,*12*). To further verify the topology of these singularities, gradient vector fields of $\Psi$ and $\Delta$ were constructed from the measured ellipsometric maps (Fig. S8). The local vector patterns reveal phase winding around the singular points, with the two ultraviolet PSs exhibiting opposite winding directions. The topological charge was then calculated by integrating the phase gradient along a closed loop enclosing each singularity, yielding opposite charges for the ultraviolet PS pair. These results confirm the zero-reflection points as topological PSs and experimentally establish the existence of an oppositely charged PS pair in the exciton enabled $MoS_2$ system.

## Thickness controlled evolution and annihilation of oppositely charged PSs

Following the identification of the oppositely charged PS pair in the $MoS_2$/sapphire system, a series of $MoS_2$ films with thicknesses ranging from 0.69 to 2.37 nm were prepared to investigate the thickness controlled evolution of the PSs. Fig. 4a shows photographs of the ten $MoS_2$ samples with different thicknesses, exhibiting a gradual increase in optical contrast with increasing film thickness. The PSs in these samples were then characterized using SE, and the extracted incidence angles and wavelengths are summarized in Fig. 4b. Two distinct regimes can be identified: a three PS regime at smaller thicknesses and a single PS regime at larger thicknesses. The two ultraviolet PSs carrying opposite topological charges near 310-330 nm approach each other with increasing film thickness and eventually annihilate at a critical thickness of approximately 1.2 nm, leaving only the singularity near 420 nm, which is associated with the excitonic resonance.

To further characterize this evolution, the incidence angles and wavelengths of the three PSs were extracted as a function of film thickness (Fig. 4c,d). Importantly, the incidence angles exhibit a stronger response to thickness variation than the corresponding wavelengths. The incidence angles of all three PSs shift to higher angles as the film thickness increases (Fig. 4c). In contrast, the wavelengths show only weak thickness dependence: the two ultraviolet PSs exhibit slight spectral shifts, whereas the singularity near 420 nm remains nearly unchanged (Fig. 4d). These thickness dependent shifts indicate that the PSs continuously evolve in the wavelength-incidence angle parameter space, and the oppositely charged ultraviolet PS pair gradually approaching each other. With further thickness increase, the pair reaches a critical annihilation point and disappears through charge conserving annihilation, leaving the excitonic PS near 420 nm (Fig. 4e).

To understand how film thickness drives the topological evolution of PSs, the dielectric responses of four representative $MoS_2$ films with thicknesses of 0.72, 1.05, 1.20 and 2.20 nm were extracted. Their structural and optical characterizations are summarized in Figs. S4 and S5. The intersections between the $MoS_2$ dispersion curves and the singular line provide a direct geometric picture of thickness induced PS evolution in the $(n,k)$ parameter space. As shown in Fig. 4f,g, the $n-k$ dispersion curves are progressively reshaped as the $MoS_2$ thickness increases. The closed loop in the ultraviolet region, as indicated by the arrow in Fig. 4f, gradually shrinks with increasing thickness causing the two intersections with the singular line to approach each other and eventually disappear. In contrast, the $n-k$ trajectory around the C excitonic region does not form a closed loop and maintains its intersection with the zero-reflection line. This thickness-controlled evolution drives the annihilation of the oppositely charged PS pair while preserving the total topological charge. At the same time, the amplitude of the E feature in the imaginary part of the dielectric function decreases with increasing thickness, indicating a reduction in the oscillator strength of the associated high energy transition. According to the Kramers–Kronig relation, the reduction of the absorptive response is accompanied by a corresponding modification of the dispersive component, which reshapes the ultraviolet $n-k$ dispersion loop.

To better resolve the thickness dependent changes of the E transition, the second derivative spectra of the real and imaginary parts of the dielectric function were fitted using a critical point (CP) model (Supplementary Note 3) (*32*, *33*). Six CP transitions were resolved, among which CP5 corresponds to the E feature (Fig. S7). With increasing $MoS_2$ thickness, CP5 shifts to lower energy and exhibits increased broadening, indicating that the E related transition becomes red shifted and more spectrally broadened (Table S2). These thickness dependent changes in CP5 provide microscopic evidence for the evolution of the dielectric response underlying the observed topological transition. Together, these results reveal that thickness induced modifications of the dielectric response reshape the ultraviolet $n-k$ dispersion loop, driving the system across the zero reflection transition boundary and leading to annihilation of the oppositely charged PS pair. To the best of our knowledge, these result demonstrate thickness controlled evolution and charge conserving annihilation of topological PSs in atomically thin $MoS_2$, providing a practical route for engineering topological optical states through intrinsic material dispersion in simple thin film platforms.

## Evaluation of the sensitivity of topological PSs

The phase sensitivity of the topological PSs was evaluated using a standard refractive index perturbation experiment, in which controlled variations of the air refractive index (RI) were generated by tuning the ambient pressure. Before the experiments, numerical simulations were performed to assess the responses of $\Psi$ and $\Delta$ to small refractive index variations, and the full results are presented in the Fig. S10. Experimentally, SE measurements were performed using a home built sealed chamber mounted on the sample stage of the ellipsometer. A 1.05 nm $MoS_2$ film on a sapphire substrate was mounted inside the chamber (Fig. S9). The gauge pressure was varied from 0 to -0.08 MPa in 0.02 MPa steps, corresponding to a decrease in the refractive index of air from 1.000293 to 1.000062 in steps of 0.000058. The corresponding refractive index of air was calculated using the standard pressure-dependent relation (details see Supplementary Information note 5) (*52*).

As predicted, near the three PSs, the change in Ψ, (Fig.5a, e, i) is relatively small, whereas Δ (Fig. 5b,f,j) shows a dramatic dependence on the refractive index of air owing to the rapid phase variation. To better visualize the phase sensitivity, the δΔ was defined as the difference between the Δ spectrum measured at each pressure condition and the reference Δ spectrum measured at ambient pressure ($n = 1.000293$). In Fig. 5c,g,k, dramatic changes in δΔ are observed near each of the three PSs in response to minute RI variations of air. Accordingly, the phase sensitivity was defined as δΔ/δn. At the three characteristic wavelengths of 321, 316 and 421 nm (marked by blue arrows), δΔ varies systematically with δn (Fig. 5d,h,l). The maximum experimental sensitivities reach $9.35\times10^4$, $8.55\times10^4$ and $7.43\times10^4$ deg/RIU for the three PSs, respectively, with the ultraviolet PS showing the highest sensitivity. Notably, the maximum phase sensitivity achieved in our device reaches $9.35 \times 10^4$ deg/RIU, which is comparable to previously reported PS based sensing systems (*24*). These results demonstrate that atomically thin $MoS_2$ films provide highly sensitive phase responses near topological PSs, highlighting their potential for compact phase sensitive RI measurements.

## DISCUSSION

Topological PSs offer a unique opportunity to explore the evolution of optical states in complex parameter spaces. Here, the evolution of PSs is experimentally demonstrated in an atomically thin $MoS_2$ system. Combined theoretical and experimental analyses reveal that the charge paired PSs are associated with the closed loop formed by the high energy dielectric dispersion of $MoS_2$ in the $n-k$ parameter space. The loop crosses the zero reflection condition with opposite orientations and gives rise to PSs with opposite topological charges. By tuning the $MoS_2$ thickness, the dielectric response is modified and the high energy $n-k$ dispersion loop is reshaped, driving the evolution and annihilation of the oppositely charged PS pair while preserving the total topological charge. This work establishes dielectric dispersion as a material parameter for manipulating topological optical states in atomically thin materials. The strong phase response near these singularities further enables sensitive RI measurements, with sensitivities comparable to previously reported PS based sensing systems. This approach may be extended to other two-dimensional semiconductors with strong excitonic and high-energy optical responses, providing a general route for exploring material controlled topological optical states in ultrathin systems.

## MATERIALS AND METHODS

### Sample Fabrication

$MoS_2$ thin films with different thicknesses were prepared by a two-step growth process consisting of Mo deposition and subsequent sulfurization. Single-crystal c-sapphire substrates, $Al_2O_3$(0001), were purchased from Nanjing MKNANO Tech. Co., Ltd. Before growth, the substrates were ultrasonic cleaned in acetone, ethanol, and deionized water for 10 min, followed by blow drying with nitrogen and oxygen plasma treatment of 50 w under vacuum for 5 min. The substrates were cleaned and annealed at 1100°C for 4 h in air to produce atomically stepped surfaces, then load into a home-built molecular beam epitaxy system with a base pressure of $1.0\times10^{-9}$ Torr and degassed at 200°C for 20 min. Ultrathin Mo layers with controlled thicknesses were deposited onto the substrates by electron beam evaporation while the substrate temperature was maintained at room temperature. During deposition, the Mo thickness was monitored in situ by differential reflectance spectroscopy during deposition. The differential reflectance spectrometer was attached to the home-built molecular beam epitaxy system through a strain free optical window, enabling reflectance spectra to be recorded continuously without interrupting the growth process. The differential reflectance signal was calculated from the reflectance spectra acquired during Mo deposition, using the reflectance spectrum of the bare c-sapphire substrate as the reference. The evolution of the signal was used to monitor the Mo growth process and guide thickness control. Details of the differential reflectance analysis are provided in the Supplementary Note 2.

$MoS_2$ thin films were synthesized by a gradient sulfur-supply method in a home-built CVD tube furnace system (*25*, *53*) equipped with independently controlled source tubes and separate carrier gas lines. Sulfur powder with a purity of 99.95% (Aladdin Inc.) was used as the sulfur source. Before sulfurization, the system was evacuated to a pressure below 10 Pa. An 80 sccm Ar flow was maintained as the background carrier gas during heating and sulfurization. The sulfur zone was heated to 200 °C over 30 min, while the substrate zone containing the Mo-coated sapphire substrates was heated to 810 °C over 60 min. Once the sulfur zone reached 200 °C, the sulfur source was moved to the designated position, and Ar carrier gas was introduced through the sulfur source line at 10 sccm. When the substrate zone reached 810 °C, the Ar flow rate through the sulfur source line was increased to 20 sccm. Sulfurization was carried out for 20 min. After sulfurization, the sulfur source was withdrawn, and the Ar flow through the sulfur source line was stopped. The background Ar flow was then increased to 200 sccm, and the system was allowed to cool slowly to room temperature under continuous pumping. All $MoS_2$ films were prepared using the same sulfurization parameters, including temperature and carrier gas flow rate. Only the thickness of the predeposited Mo layer was precisely varied to obtain $MoS_2$ films with different thicknesses.

## Atomic force microscopy

The thickness and surface morphology of the $MoS_2$ films were characterized using a Bruker Dimension Icon atomic force microscope operated in tapping mode under ambient conditions.

## Scanning electron microscopy

The images were acquired using a Thermo Fisher Scientific Apreo S scanning electron microscope operated in standard mode at 5 kV and 0.1 nA under a vacuum of less than $1\times10^{-3}$ Pa.

## Raman and PL measurement

Raman and PL spectra were collected using a Renishaw inVia confocal microscope. All measurements were performed with a 532 nm excitation laser, a 50× objective lens, and a 2400 grooves $mm^{-1}$ grating.

## Ellipsometry measurements

PSs were visualized using a variable angle spectroscopic ellipsometer (J.A. Woollam RC2-XI) equipped with a 75 W xenon arc lamp, covering a spectral range of 210-1690 nm and an incidence angle range of 45-90°. Measurements were carried out in the standard angular scanning mode at room temperature in air, with spectral and angular resolutions of 1 nm and 0.02°, respectively. $\Psi$ and $\Delta$ were recorded as functions of wavelength ($\lambda$) and incidence angle ($\theta$), yielding two-dimensional $\Psi$ and $\Delta$ maps. The recorded values were limited to the range from $-\pi/2$ to $3\pi/2$ by the periodic arctangent function. During data processing, values below $-\pi/2$ were unwrapped by adding $2\pi$ to facilitate visualization of the PSs.

## Measurement of phase sensitivity to RI variations

A home-built sealed chamber was placed on the sample stage of the ellipsometer to evaluate the response of the PSs to pressure-induced changes in the RI of air. The chamber was connected to a vacuum pump and a pressure gauge, enabling the gauge pressure inside the chamber to be adjusted from 0 to -0.08 MPa and monitored in real time. Ellipsometric spectra were recorded at different pressures, and the corresponding shifts in the singularity-related features were analyzed to evaluate the sensitivity to changes in the RI.

## Acknowledgements

**Funding:** This work was supported by National Natural Science Foundation of China (52475566, 62205241); the Metrology Science and Technology Program of Tianjin (2024TJMT008); Tianjin Science and Technology Project (24ZXZSSS00400).

**Author contributions:** YJ.W. and YH.W. contributed equally to this work. W.S. conceived and directed the project. YJ.W. and YH.W. performed the theoretical simulations and experiments and analyzed the data. G.F. and Q.L. performed theoretical simulations and contributed to data analysis. Y.L. and L.S. provided theoretical support and contributed to data analysis. C.H. supervised the project. YJ.W. wrote the original manuscript. C.H., W.S., and all authors reviewed and edited the manuscript. All authors discussed the results and commented on the manuscript.

**Competing interests:** The authors declare no competing financial interest.

**Data Availability:** All relevant data supporting the findings of this study, including the raw spectroscopic ellipsometry data and additional datasets generated and/or analyzed during the current study, are available from the corresponding authors upon reasonable request.

## Figures and Tables

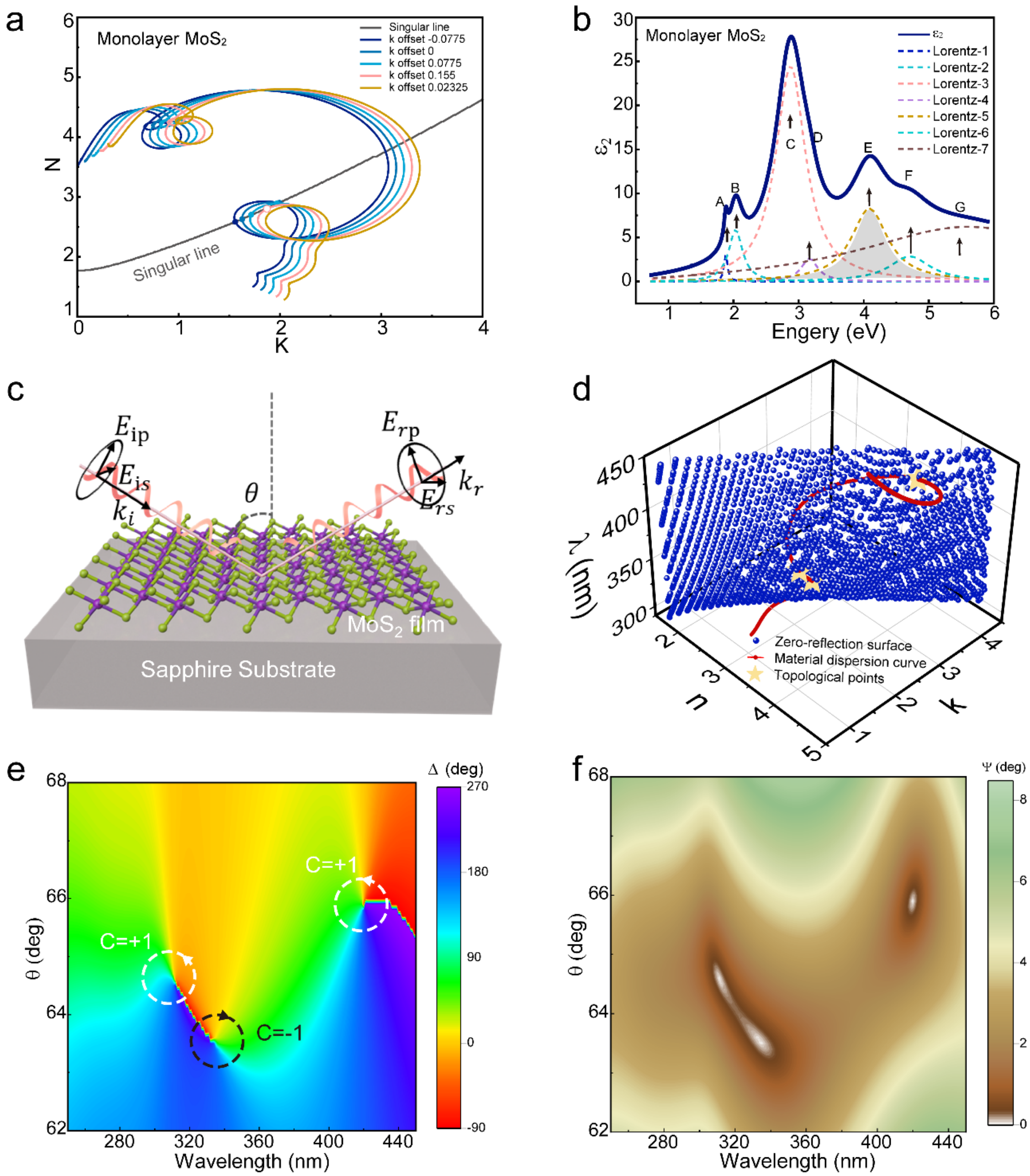

**Fig. 1 Theoretical prediction of oppositely charged PS pairs.** (a) Calculated $n-k$ dispersion curve of monolayer $MoS_2$ with different uniform offsets applied to $k$. The grey curve denotes the calculated singular line. (b) The imaginary part of the dielectric function, $\varepsilon_2$, of the $MoS_2$ film extracted from Spectroscopic Ellipsometry (SE) measurements and fitted using multiple Lorentz oscillators. (c) Schematic of the reflection geometry used to measure topological PSs in a $MoS_2$ film on a sapphire substrate with roughened

backside. (d) The $MoS_2$ dispersion curve (red solid line) intersects the ZRS, giving rise to three zero-reflection points (yellow stars). (e) Simulated maps of the ellipsometric parameter Δ. In the vicinity of the zero-reflection points, topological charges of $C = +1$ and $C = -1$ correspond to phase winding of $+2\pi$ or $-2\pi$, respectively. (f) Simulated maps of the ellipsometric parameter Ψ, from which the three zero-reflection points can be identified.

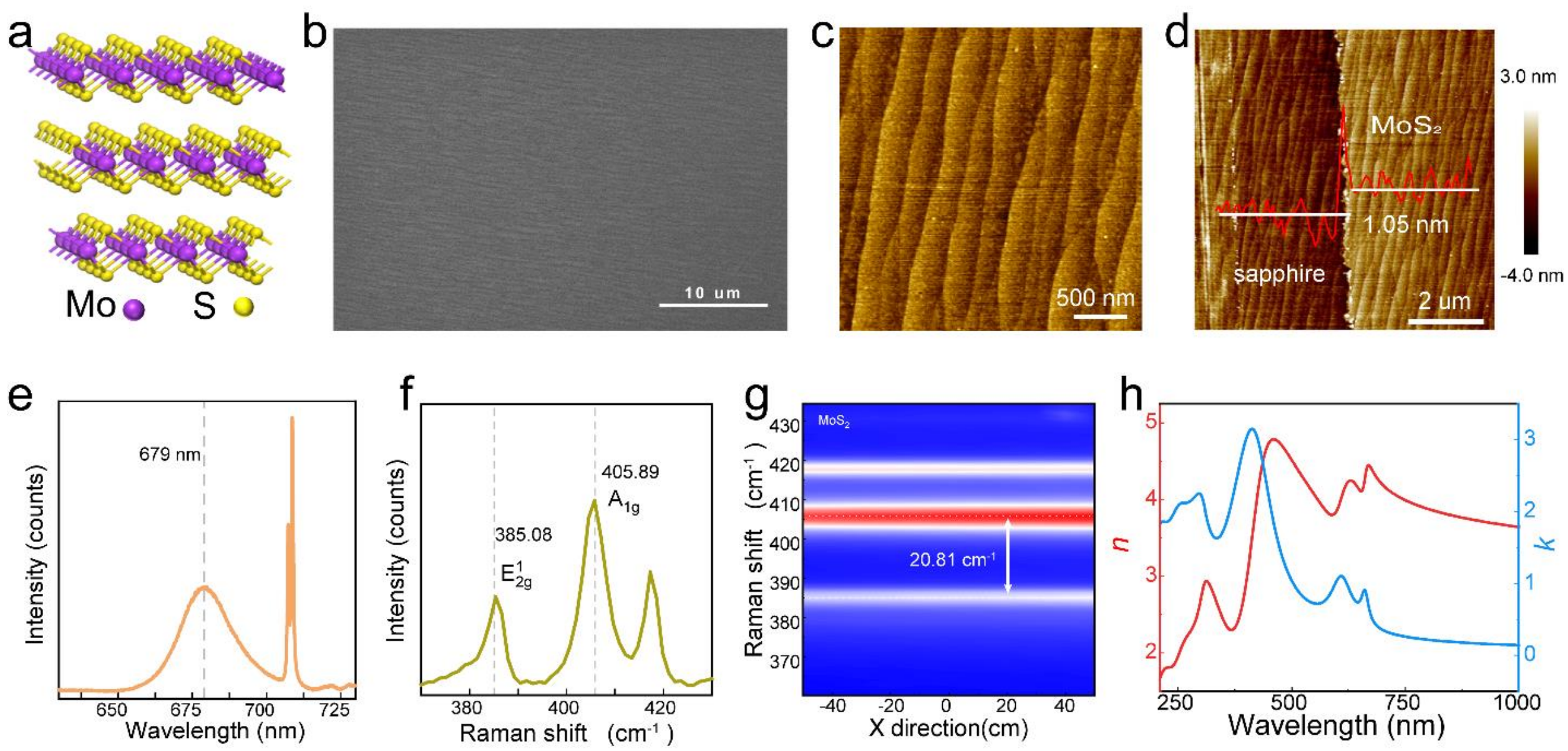

**Fig. 2 Characterization of the $MoS_2$ film.** (a) Schematic illustration of the crystal structure of $MoS_2$. (b) SEM image of the $MoS_2$ film. (c), (d) AFM topography of the $MoS_2$ film on a sapphire substrate, showing a film thickness of 1.05 nm. (e), (f) PL and Raman spectra of the $MoS_2$ film measured with a 532 nm excitation laser. (g) Spatial map of the peak separation between the $E_{2g}^{1}$ and $A_{1g}$ Raman modes of the $MoS_2$ film. (h) Optical constants of the $MoS_2$ film over the spectral range from 210 to 1000 nm. The optical model is described in Supplementary Note 3.

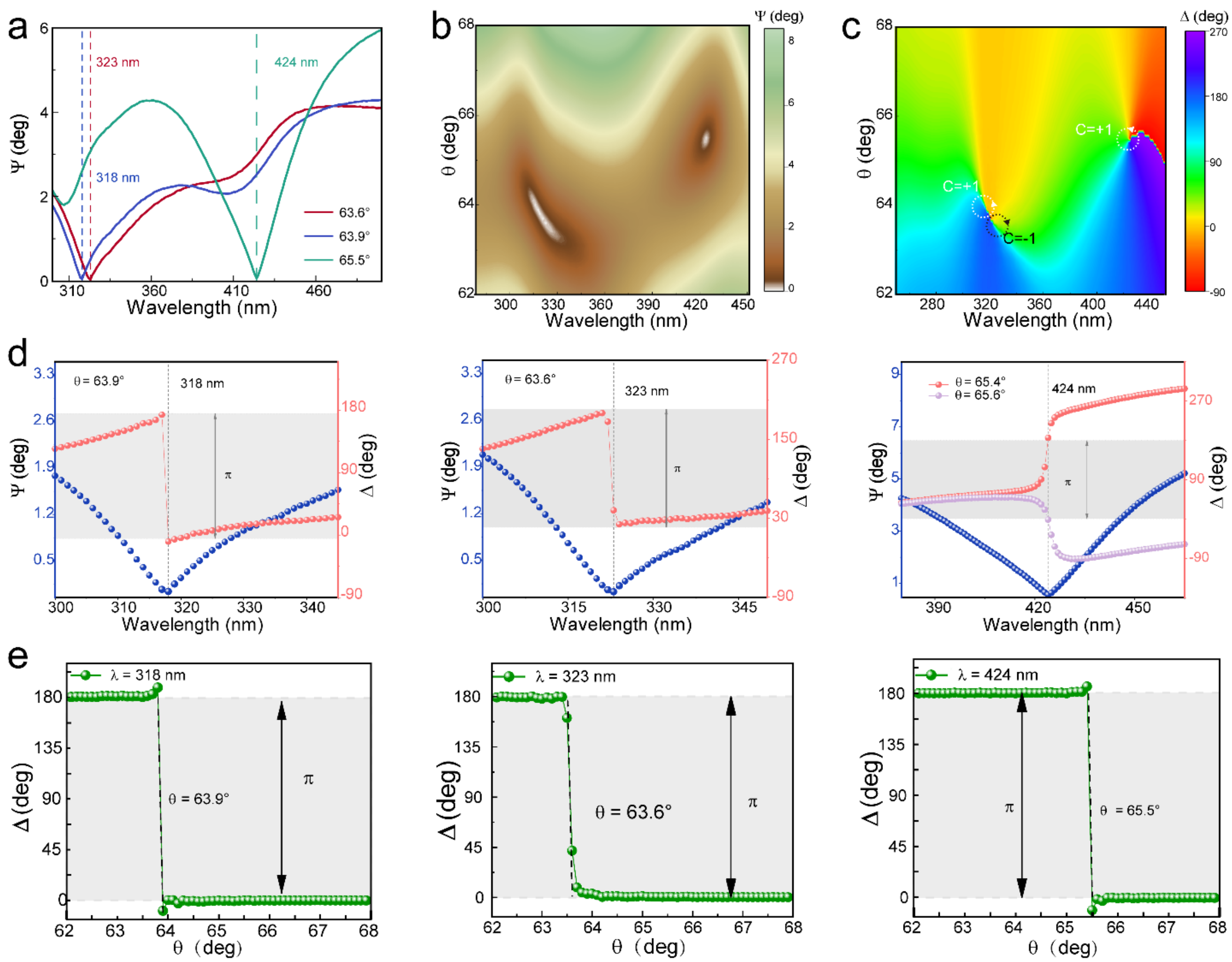

**Fig. 3 Measurement of topological PSs in a $MoS_2$ film on sapphire.** (a) Experimental Ψ spectrum, showing zero-reflection points at the indicated incidence angle. (b) Experimental Ψ map plotted as a function of wavelength and incidence angle, identifying the locations of the PSs. (c) Experimental Δ map revealing the singular phase behavior in the vicinity of zero-reflection points. The calculated topological charges are marked in the Δ map: the white dashed circles indicate the $C = +1$ singularities, whereas the black dashed circle indicates the $C = -1$. (d) Extracted Ψ and Δ spectra from panels (b) and (c) at incidence angles of 63.9° and 63.6°, as well as slightly below (65.4°) and above (65.6°) the PS condition at 65.5° ± 0.1°. (e) Extracted Δ spectrum as a function of incidence angle at wavelengths of 318, 324, and 424 nm, showing that Δ exhibits an abrupt $\pi$ phase jump upon crossing the PS scondition.

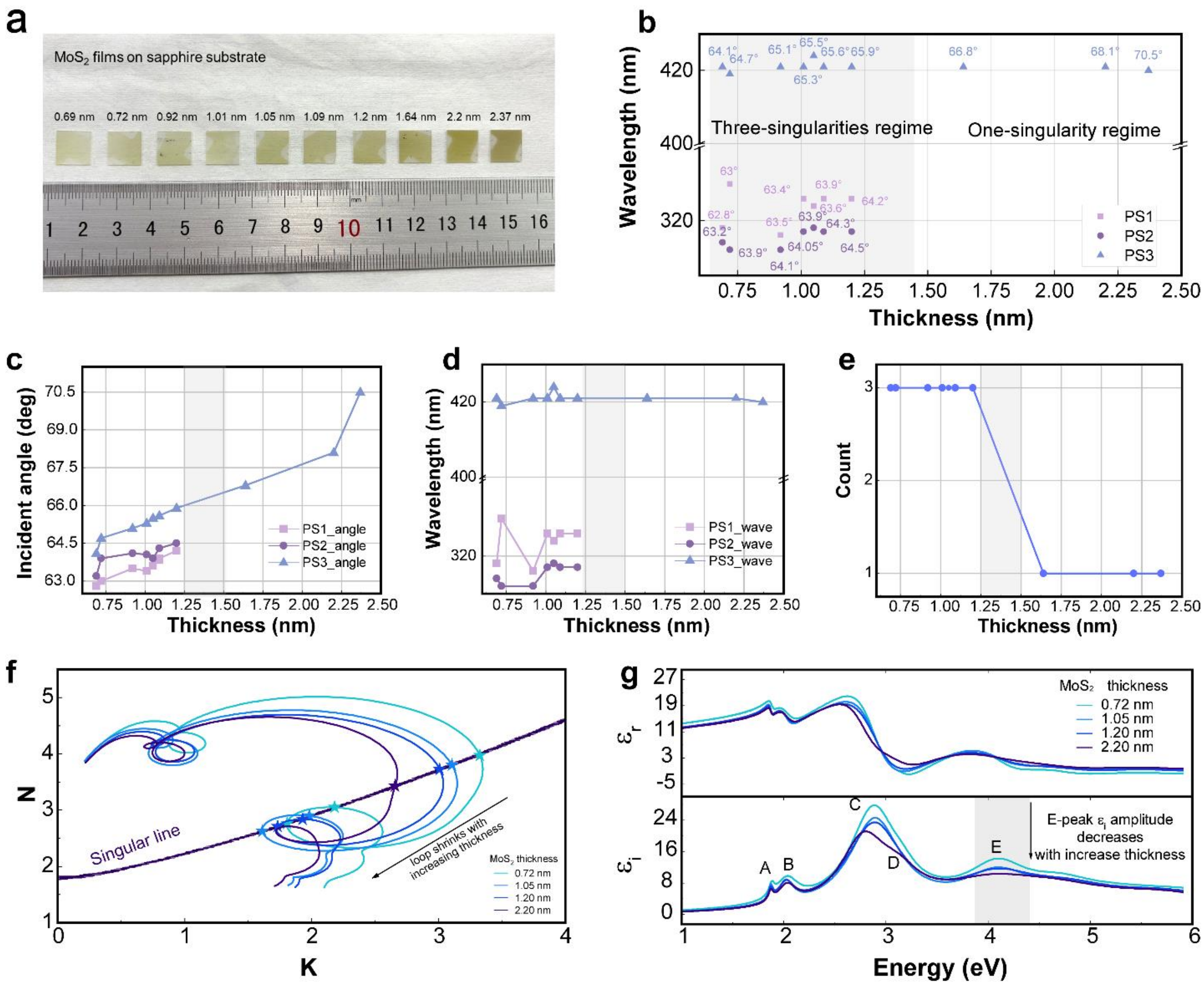

**Fig. 4 Thickness controlled evolution of PSs in $MoS_2$ films on sapphire.** (a) Photograph of $MoS_2$ films with different thicknesses on sapphire substrates. (b) Summary of the phase-singularity conditions obtained from ellipsometric measurements, showing the corresponding wavelengths and incidence angles of PS1, PS2, and PS3, which are denoted by squares, circles, and triangles respectively. The shaded region marks the transition from the three PSs regime to the single PS regime. (c,d) Thickness controlled evolution of the incidence angles and wavelengths corresponding to the three PSs. (e) Thickness controlled evolution of the number of PSs. (f) Dispersion curves of $MoS_2$ films with different thicknesses and the singular line in the $n - k$ plane. Stars indicate the PSs at the intersections of the dispersion curves and the singular line. (g) Dielectric function of $MoS_2$ films with different thicknesses.

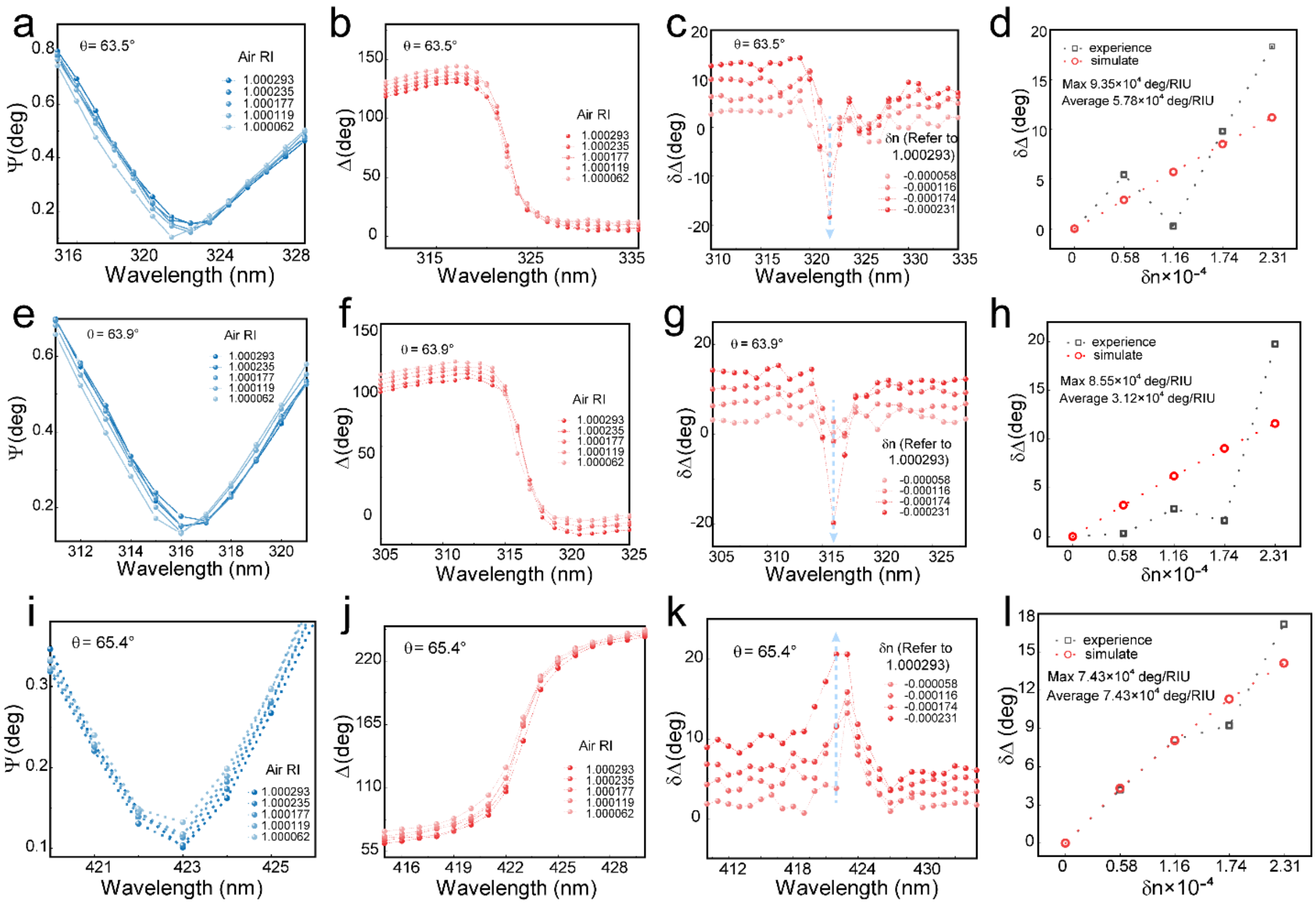

**Fig. 5 Evaluation of RI sensitivity of phase singularities.** (a,b,e,f,i,j) Variations of the ellipsometric parameters Ψ and Δ with air RI at the incidence angles corresponding to PS1, PS2, and PS3, respectively. (c,g,k) Differential spectra δΔ calculated relative to the reference air RI at room temperature and atmospheric pressure (n=1.000293). (d,h,l) Extracted RI sensitivities of the three PSs. The red dashed lines denote the simulated sensitivities, and the black dashed lines denote the experimentally measured sensitivities.

# Supplementary Materials for

## Charge-conserving creation and annihilation of topological phase singularities in atomically thin semiconductors

Yongjing Wang[#], Yuhang Wang[#] *et al.*

*Corresponding author. cghu@tju.edu.cn; wfshen@tju.edu.cn.

**This PDF file includes:**

## Supplementary Note 1: Optical model and prediction of topological phase singularities in $MoS_2$ films

To predict the topological phase singularities (PSs) in $MoS_2$ films on sapphire substrates, a three-layer optical model was constructed, consisting of air, $MoS_2$ film and sapphire substrate. Because the backside of the sapphire substrate was roughened, back-reflected light from the rear surface was suppressed, allowing the sapphire substrate to be treated as a semi-infinite medium in the optical calculation. The optical response of the air/$MoS_2$/sapphire structure was calculated using the transfer-matrix method (*38*). For each layer $m$, the complex refractive index was defined as:

$$N_m(\lambda) = n_m(\lambda) - ik_m(\lambda) \tag{1}$$

where λ is the wavelength of the incident light, $n$ is the refractive index and $k$ is the extinction coefficient. In the three-layer model, $m = 0$, $m = 1$ and $m = 2$ correspond to air, the $MoS_2$ film and the sapphire substrate, respectively. The refraction angle in each layer was calculated using Snell's law:

$$N_{m-1} sin\theta_{m-1} = N_m sin\theta_m \tag{2}$$

where $\theta_m$ is the propagation angle in layer $m$. The Fresnel reflection coefficients at the interface between layers $m-1$ and $m$ are given by

$$r_m^p = \frac{N_m cos\theta_{m-1} - N_{m-1} cos\theta_m}{N_m cos\theta_{m-1} + N_{m-1} cos\theta_m} \tag{3}$$

$$r_m^s = \frac{N_{m-1} cos\theta_{m-1} - N_m cos\theta_m}{N_{m-1} cos\theta_{m-1} + N_m cos\theta_m} \tag{4}$$

For the $MoS_2$ film with thickness $d$, the propagation phase is given by

$$\beta = \frac{2\pi}{\lambda} N_1 \, d \, cos\theta_1 \tag{5}$$

For each polarization, the total transfer matrix was constructed by multiplying the interface and propagation matrices across the air/$MoS_2$/sapphire structure:

$$M^{p,s} = \prod_m I_m^{p,s} P_m \tag{6}$$

where $I_m^{p,s}$ is the interface matrix and $P_m$ is the propagation matrix of layer $m$. The interface matrix is given by

$$I_m^{p,s} = \frac{1}{t_m^{p,s}} \begin{pmatrix} 1 & r_m^{p,s} \\ r_m^{p,s} & 1 \end{pmatrix} \tag{7}$$

and the propagation matrix is

$$P_m = \begin{pmatrix} e^{i\beta_m} & 0 \\ 0 & e^{-i\beta_m} \end{pmatrix} \tag{8}$$

With

$$\beta_m = \frac{2\pi}{\lambda} N_m \, d_m \, cos\theta_m \tag{9}$$

The total reflection coefficients were then obtained from the elements of the total transfer matrix as:

$$r^s = \frac{M_{21}^s}{M_{11}^s} \tag{10}$$

$$r^p = \frac{M_{21}^p}{M_{11}^p} \tag{11}$$

The ellipsometry ratio was calculated from the total p- and s-polarized reflection coefficients as

$$\rho = \frac{r_p}{r_s} = tan\Psi e^{i\Delta} \tag{12}$$

here $\Psi$ represents the amplitude ratio and $\Delta$ is the phase difference between the p and s polarized reflected fields. The p-polarized zero-reflection condition corresponds to $r^p = 0$, for which $\Psi$ approaches zero and $\Delta$ becomes singular.

Using the calculated reflection coefficients, the zero-reflection condition was searched in the combined $(n, k, \theta, \lambda)$ parameter space. Specifically, $n$ and $k$ of the $MoS_2$ layer were swept as optical-constant variables, and for each $(n, k)$ pair, the wavelength $\lambda$ and incidence angle $\theta$ were scanned to find the minimum value of $\Psi$. Points satisfying $\Psi < 0.01°$ were recorded as zero-reflection points, which form the zero-reflection surface (ZRS) in the $(n, k, \theta, \lambda)$ parameter space. In the calculation, $n$ was varied from 1 to 5, $k$ from 0 to 4, $\lambda$ from 300 to 500 nm, and $\theta$ from $60°$ to $68°$, with step sizes of 0.01 for both $n$ and $k$, 1 nm for $\lambda$, and $0.1°$ for $\theta$. The $MoS_2$ film thickness was set according to the experimentally extracted value. The extracted $MoS_2$ dispersion curve, $n(\lambda)$ and $k(\lambda)$, obtained from spectroscopic ellipsometry (SE) fitting as described in Supplementary Note 3, was then superimposed onto the calculated ZRS. According to the Jordan–Brouwer separation theorem (*54*, *55*), when the $MoS_2$ dispersion curve starts and ends on opposite sides of the ZRS, as shown in Fig. 1b, the curve must intersect the ZRS, thereby providing zero-reflection points. Small imperfections in $MoS_2$ may slightly modify its optical constants and shift the dispersion curve, but the curve still crosses the ZRS as long as it starts and ends on opposite sides. This indicates that the zero-reflection points are topologically protected by the Jordan–Brouwer theorem.

From another perspective, the PSs can be described in the $(n, k)$ parameter plane by fixing $\lambda$ and $\theta$. For each given $(\lambda, \theta)$, $n$ and $k$ were scanned as independent variables to calculate $\Psi$ and $\Delta$. The zero-reflection condition forms a singular line in the $(n, k)$ plane, across which $\Delta$ undergoes a Heaviside-like $\pi$-phase jump (*12*). A real phase singularity occurs when the experimentally extracted $MoS_2$ dispersion curve, intersects this singular line. Thus, the intersection between the singular line and the material dispersion curve provides a geometric criterion for identifying and tracking PSs.

## Supplementary Note 2: Synthesis and morphological characterization of $MoS_2$ films

$MoS_2$ films with different thicknesses were prepared using a two-step growth strategy, as schematically illustrated in Fig. S2. First, Mo precursor films with controlled thicknesses were deposited on c-plane sapphire substrates by molecular beam epitaxy. During Mo deposition, an assembled DRS system was used for in-situ optical monitoring, allowing the deposition process to be followed in real time. By precisely controlling the Mo deposition time, precursor films with different thicknesses were obtained on sapphire substrates. The deposited Mo films were then converted into $MoS_2$ films through sulfurization in a home-build chemical vapor deposition (CVD) system (*25*). The system consists of a three-zone CVD furnace and three independently movable quartz sub-tubes driven by separate electric motors, which allows the sulfur source position to be precisely controlled during the sulfurization process. By controlling the Mo deposition time before sulfurization, this two-step process enables the preparation of $MoS_2$ films with different thicknesses under the same sulfurization conditions.

The annealed sapphire substrate was first examined by atomic force microscopy (AFM) before Mo deposition. As shown in Fig. S3a, the surface consists of well-defined step-and-terrace features, with an average terrace width of approximately 143 nm and a step height of approximately 4 Å. Such a regular stepped morphology indicates that the sapphire surface is atomically smooth and sufficiently ordered for the deposition of ultrathin Mo precursor films. After 90 min of Mo deposition, the stepped morphology remains visible over a larger scan area (Fig. S3b). No obvious island-like aggregates or severe surface roughening are observed, indicating that the Mo deposition proceeds without destroying the underlying stepped sapphire morphology.

After sulfurization, continuous $MoS_2$ films with different thicknesses were obtained. Large-area AFM images of the four representatives $MoS_2$ films used in Fig. 4, with thicknesses of 0.72, 1.05, 1.20, and 2.20 nm, are shown in Fig. S4. The films show continuous surface coverage over the measured regions, confirm the controllable preparation of ultrathin $MoS_2$ films with different thicknesses. Raman spectroscopy was used to confirm the formation of $MoS_2$ and to assess the thickness-dependent structural response of the films. As shown in Fig. S5a, all samples exhibit the characteristic in-plane $E_{2g}^{1}$ and out-of-plane $A_{1g}$ modes of $MoS_2$ near 385 and 405 $\mathrm{cm}^{-1}$, respectively, confirming the successful conversion of the Mo precursor into $MoS_2$ after sulfurization. With the film thickness increases, both Raman modes intensify and their frequency separation becomes larger, in line with the expected layer-dependent Raman behavior of $MoS_2$ (*56*, *57*).

The corresponding Photoluminescence (PL) spectra of the same samples are shown in Fig. S5b. The emission centered near 685 nm corresponds to a bandgap of approximately 1.82 eV. A clear quenching of the PL signal is observed for thicker $MoS_2$ films, which is consistent with the transition from an ultrathin direct-gap-like response towards a more indirect-bandgap-like electronic structure, together with increased non-radiative recombination (*56*, *58*). Notably, the 0.72 nm film also exhibits relatively weak PL emission.

This is likely associated with incomplete film coverage at the ultrathin limit, where residual nanoscale gaps and edge-rich regions reduce the effective emitting area and may introduce additional nonradiative recombination pathways (*59*, *60*). Together with the AFM results, the Raman and PL measurements demonstrate that ultrathin $MoS_2$ films with controlled thicknesses were successfully prepared on sapphire substrates.

## Supplementary Note 3: Extraction of optical constants of $MoS_2$ from spectroscopic ellipsometry measurements

The optical constants of the $MoS_2$ films were extracted from SE measurements. The parameters $\Psi$ and $\Delta$ were measured over a broad spectral range of 200-1690 nm at incidence angles of $55°$, $60°$, and $65°$. Taking the 0.72 nm $MoS_2$ film as a representative example, the measured spectra and fitting results are shown in Fig. S6. The dielectric response of $MoS_2$ was modeled using a Lorentz-oscillator model that satisfies the Kramers–Kronig relations, ensuring physically consistent real and imaginary parts of the dielectric function (*26*). To physically retrieve the dielectric response of $MoS_2$, seven Lorentz oscillators were used together with a baseline constant, with the corresponding oscillator parameters summarized in Table S1. The oscillator decomposition captures the main absorption features of $MoS_2$, including A, B and C excitonic resonances and higher energy interband transitions (*45*). The model was fitted to the measured SE spectra, giving a mean squared error (MSE) of 1.54, indicating good agreement between the calculated and experimental data. The resulting total $k$ spectrum shows pronounced absorption features over the visible–ultraviolet spectral region, reflecting the strong dispersive optical response of the film. The complex dielectric function is given by Eq. (13):

$$\varepsilon(E) = \varepsilon_\infty + \sum_{j=1}^{7} \frac{A_j \Gamma_j E_j}{E_j^2 - E^2 + i\Gamma_j E} \tag{13}$$

where $E$ is the photon energy, $\varepsilon_\infty$ is the baseline dielectric constant, and $A_j$, $E_j$, and $\Gamma_j$ denote the amplitude, resonance energy, and broadening parameter of the $j$-th Lorentz oscillator, respectively. The extracted optical constants were subsequently used in the transfer-matrix calculation and ZRS analysis described in Supplementary Note 1.

It is just not accurate to extract the energy critical point by simply locating the energy at which the maximum of absorption spectrum occurs (*45*). To better resolve the overlapping CP structures, second-derivative spectra, $\frac{d^2\varepsilon}{dE^2}$ were calculated numerically as shown in Fig.S7. The linear filtering algorithm of Savitzky and Golay was simultaneously applied to minimize noise amplification and lineshape distortion. The second derivative spectra of the dielectric function were fitted using a critical point model:

$$\frac{d^2\varepsilon}{dE^2} = n(n-1)A_{amp}\, e^{i\phi}\left(E - E_g + i\Gamma\right)^{n-2}, n \neq 0 \tag{14}$$

$$\frac{d^2\varepsilon}{dE^2} = A_{amp}\, e^{i\phi}\left(E - E_g + i\Gamma\right)^{-2}, n = 0 \tag{15}$$

where $A_{amp}$ is the amplitude, $E$ is the threshold energy, $\Gamma$ is the broadening, and $\phi$ is the phase. The exponent $n$ has the values -1, -1/2, 0 and 1/2 for excitonic, one-, two (logarithmic), and three-dimensional CPs. Those are adjustable parameters. The real and imaginary parts were both fitted simultaneously. All the CPs of the $MoS_2$ film are best represented with the excitonic lineshape ($n = -1$), as previously reported (*33*). The CP energies and broadening from the fits for $MoS_2$ films with different thicknesses are listed in Table S2 and Table S3.

## Supplementary Note 4: Experimental results of phase singularities.

To further verify the singular nature of the three experimentally observed zero-reflection points, gradient vector fields were constructed from the measured ellipsometric $\Psi$ and $\Delta$ maps in the wavelength and incidence-angle parameter space. As shown in Fig. S6a–c, the gradient fields of $\Psi$ point to the local zero-reflection points, identifying the amplitude minima associated with the three PSs. The corresponding $\Delta$ gradient fields show circulating vector patterns around the same positions (Fig. S6d–f), revealing the phase winding of the ellipsometric phase. The topological charge was further determined by integrating the phase gradient of $\Delta$ along closed paths enclosing the singularities, as indicated by the red dashed circles and arrows. The resulting charges are $C = +1$, $C = -1$, and $C = +1$, respectively. The coincidence of the local $\Psi$ minima and the phase winding in $\Delta$ confirms the existence of the three topological PSs.

## Supplementary Note 5: Experimental setup and calculation of pressure-dependent air refractive index

Measurements under different air refractive indices were performed using a sealed pressure-controlled chamber coupled with the spectroscopic ellipsometry system. The $MoS_2$/sapphire sample was fixed inside the chamber using double-sided tape. The chamber pressure was controlled by a pump and monitored using a pressure gauge. Quartz windows were used as the incident and exit optical windows to reduce additional optical interference during measurements. The incident light passed through the quartz window and was focused onto the $MoS_2$ film, while the reflected light was collected through the exit window by the ellipsometer. A photograph and schematic illustration of the measurement setup are shown in Fig. S9. The air refractive index at different chamber pressures was calculated using the Gladstone–Dale relation, which the gas refractivity is proportional to gas density (*52*). Since the measurements were conducted at a constant room temperature, the refractive index of air at standard atmospheric pressure was used as a reference. The refractive index under different chamber pressures was then obtained by scaling the air refractivity with the corresponding absolute pressure. After converting the gauge pressure $P_g$ to the absolute pressure $P_{\text{abs}} = P_{\text{atm}} + P_g$, the refractive index at a given chamber pressure was calculated as

$$n_{\text{air}}(P_g) = 1 + \frac{P_{\text{abs}}}{P_{\text{atm}}}[n_{\text{air}}(P_{\text{atm}}) - 1] \quad (14)$$

where $P_g$ is the gauge pressure and $P_{\text{atm}}$ is the atmospheric pressure.

## Supplementary Note 6. Numerical simulations of the sensitivity of topological PSs to refractive-index variations

Numerical simulations were first performed to evaluate how the singular phase response changes with the pressure-dependent refractive index of air. Using the $MoS_2$ optical constants $n(\lambda)$ and $k(\lambda)$ obtained from ellipsometry fitting (Supplementary Note 3), the $\Psi$ and $\Delta$ spectra were calculated for different air refractive indices following the procedure described in Supplementary Note 1. Air refractive indices corresponding to gauge pressures from 0 to $-0.08$ MPa (step 0.02 MPa) were used in the simulations. The singularities were then identified from the calculated spectra. The simulated singularities show small differences in incidence angle and wavelength relative to the experimental measurements, which may arise from experimental uncertainties such as sample alignment and the optical effects of the sealed chamber. As shown in Fig. S10, the three columns correspond to the three selected PSs. For each PS, the simulated $\Psi$ spectra are shown in Fig. S10(a,e,i), and the corresponding $\Delta$ spectra are shown in Fig. S10(b,f,j). As the air pressure decreases, the corresponding refractive index also decreases, resulting in a measurable shift of the singular phase response near the PS. To quantify this response, the phase shifts $\delta\Delta$ was calculated relative to the value at 0 MPa gauge pressure and plotted as a function of pressure in Fig. S10(c,g,k). For each pressure step from 0 to $-0.08$ MPa, the corresponding refractive-index decrement is $\delta n = 5.8 \times 10^{-5}$. Phase sensitivity was calculated as $\delta\Delta/\delta n$ (*14, 18*), and the resulting sensitivities are shown in Fig. S10(d,h,l). These simulations show that all three PSs provide a sensitive phase response to pressure-induced refractive-index variations, with phase sensitivities reaching the order of $10^4$ deg/RIU and a maximum value of $7.42 \times 10^4$ deg/RIU. This result confirms that the sharp phase variation near the zero-reflection point can be effectively used for pressure-calibrated air refractive-index measurement.

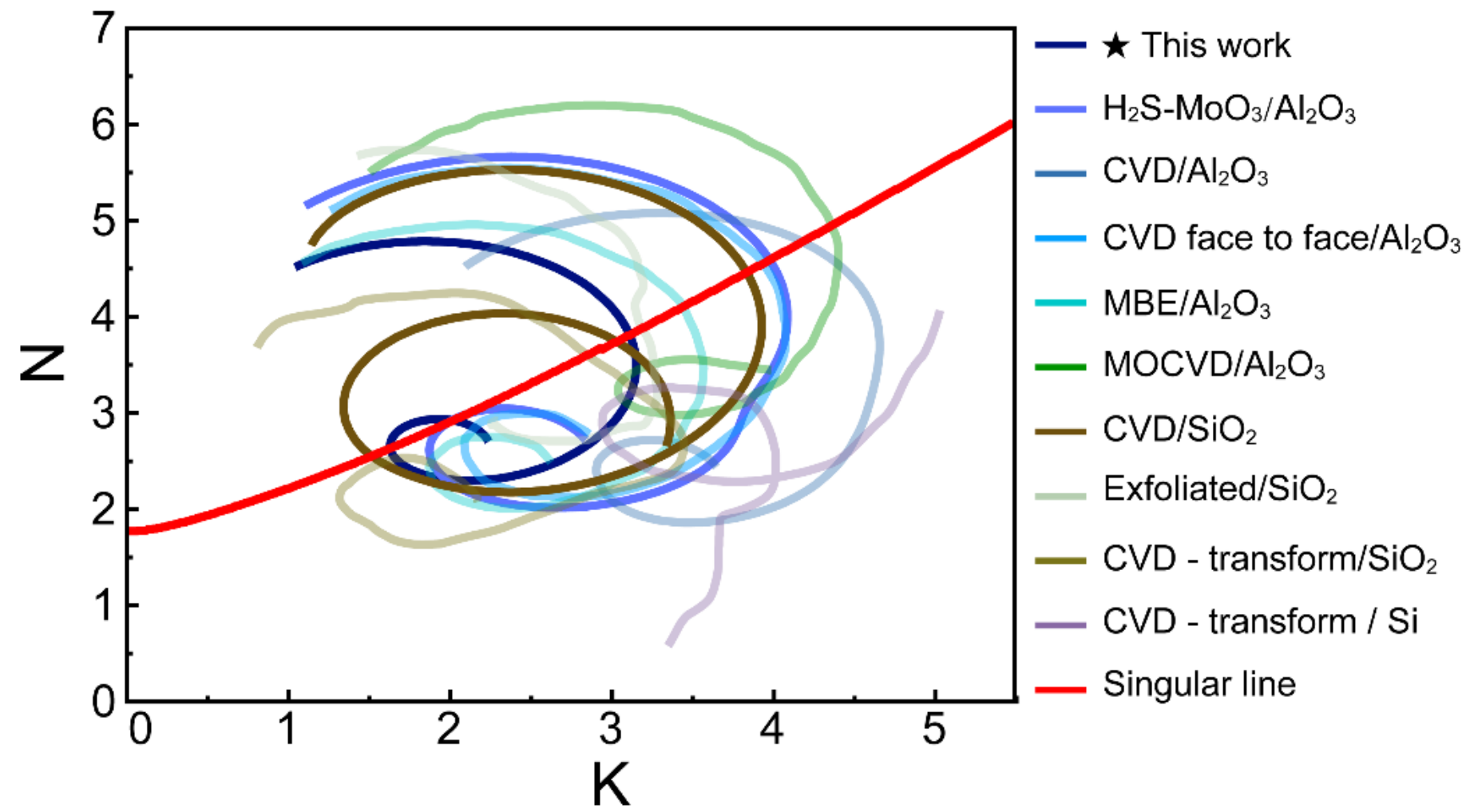

**Fig. S1. Comparison of reported monolayer $MoS_2$ $n-k$ dispersions with the singular line.**

Using this geometric criterion, we compared the $n-k$ dispersion curve of the $MoS_2$ film used in this work with reported monolayer $MoS_2$ optical constant datasets prepared under different growth, transfer and substrate conditions (Fig. S1). Although these datasets show similar overall dispersion trends, clear variations appear in the dielectric amplitude and high energy optical features. Most curves intersect the calculated singular line near the C excitonic region, whereas only a subset of the high energy $n-k$ loops near the E feature crosses the singular line. This comparison shows that the visible PS is commonly supported by the C excitonic response, while the emergence of the additional ultraviolet PS pair is more sensitive to the high energy dielectric dispersion of monolayer $MoS_2$.

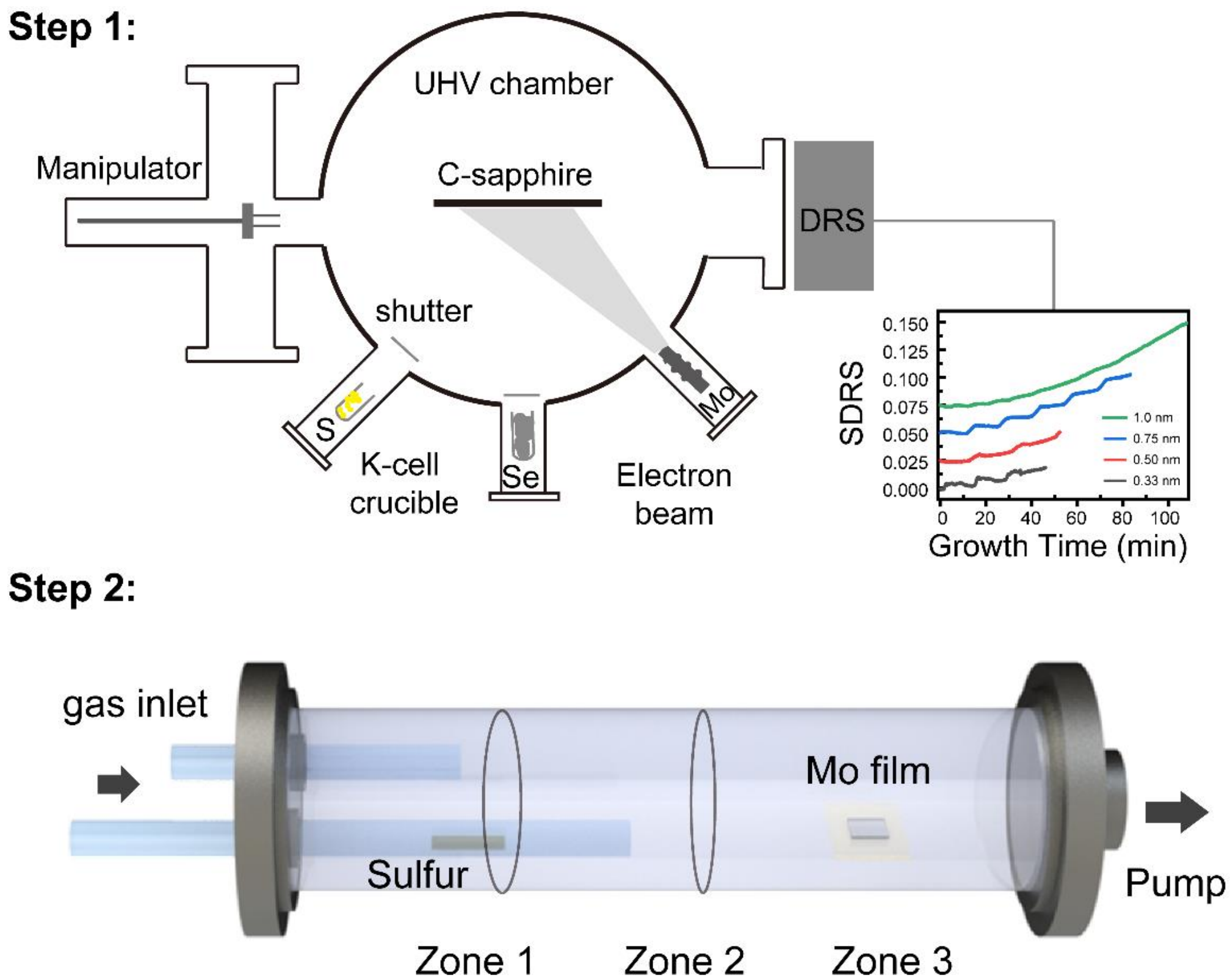

**Fig. S2. Schematic illustration of the two-step growth strategy for $MoS_2$ films on sapphire substrates.** Step 1: Mo precursor films with controlled nominal thicknesses were deposited on c-plane sapphire substrates in an ultrahigh-vacuum molecular beam epitaxy system. The deposition process was monitored in situ by differential reflectance spectroscopy (DRS), allowing the film growth to be tracked in real time. Step 2: The as-deposited Mo films were subsequently sulfurized in a home-build CVD system. The position of the sulfur source was precisely controlled to regulate the sulfur vapor supply, which then reacted with the Mo films in the growth zone to form $MoS_2$ films on sapphire substrates.

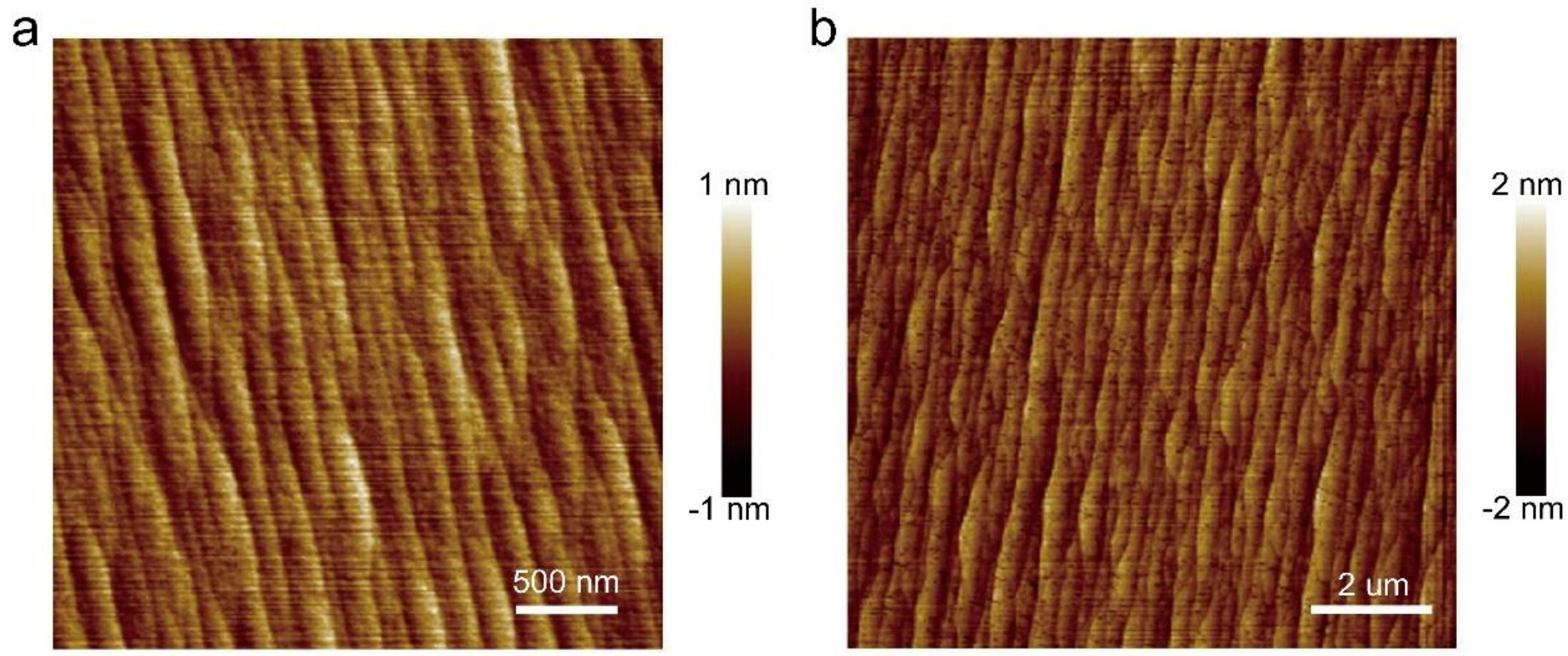

**Fig. S3. AFM characterization of sapphire substrates before and after Mo deposition.** (a) AFM image of the sapphire substrate after annealing at 1100 °C for 4 h. (b) AFM image of the sapphire substrate after deposition of Mo for 90 minutes.

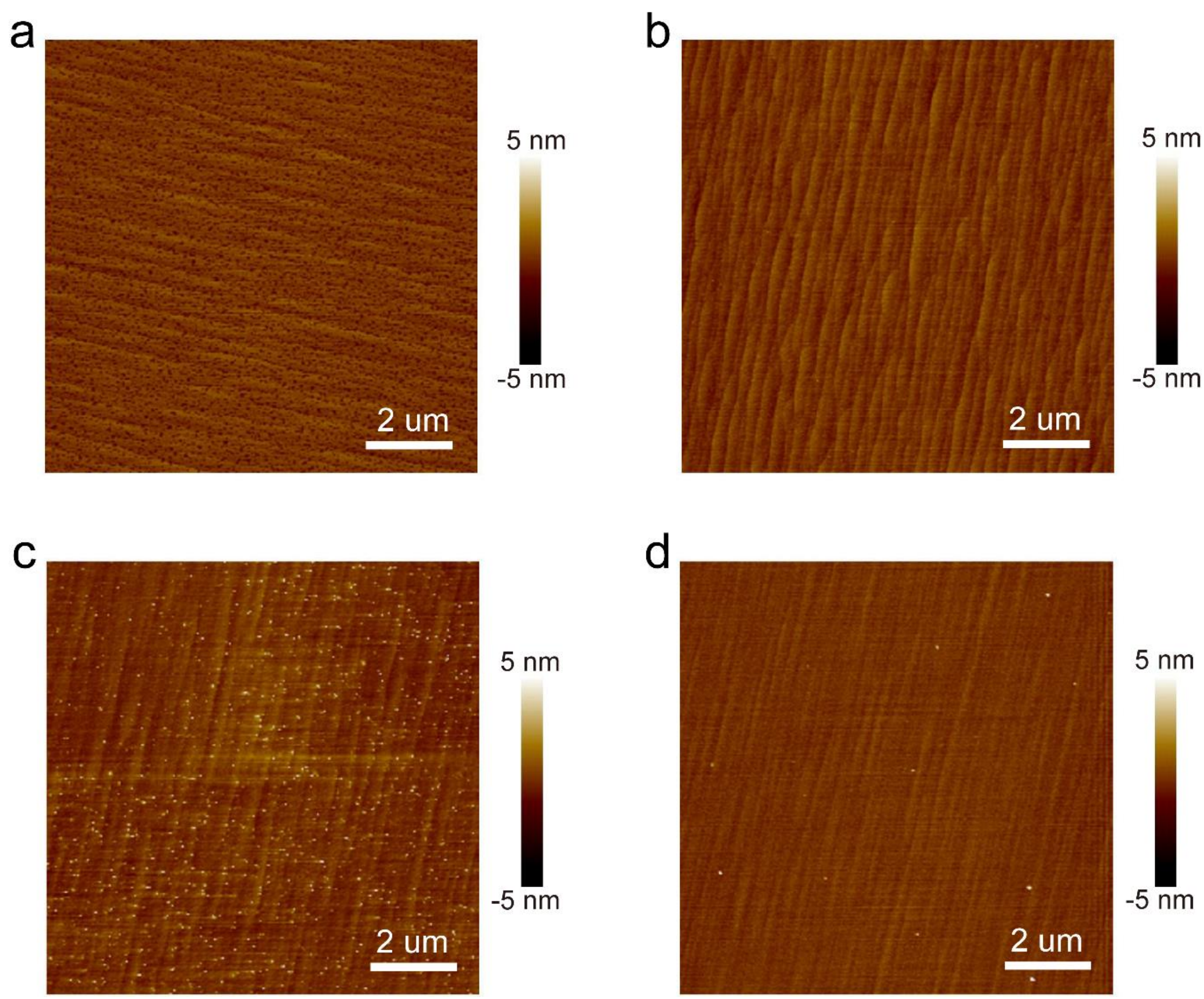

**Fig. S4. AFM characterization of $MoS_2$ films with different thicknesses.** (a) Large-area AFM images of $MoS_2$ films with thicknesses of 0.72 nm. (b) Large-area AFM images of $MoS_2$ films with thicknesses of 1.05 nm. (c) Large-area AFM images of $MoS_2$ films with thicknesses of 1.20 nm. (d) Large-area AFM images of $MoS_2$ films with thicknesses of 2.20 nm.

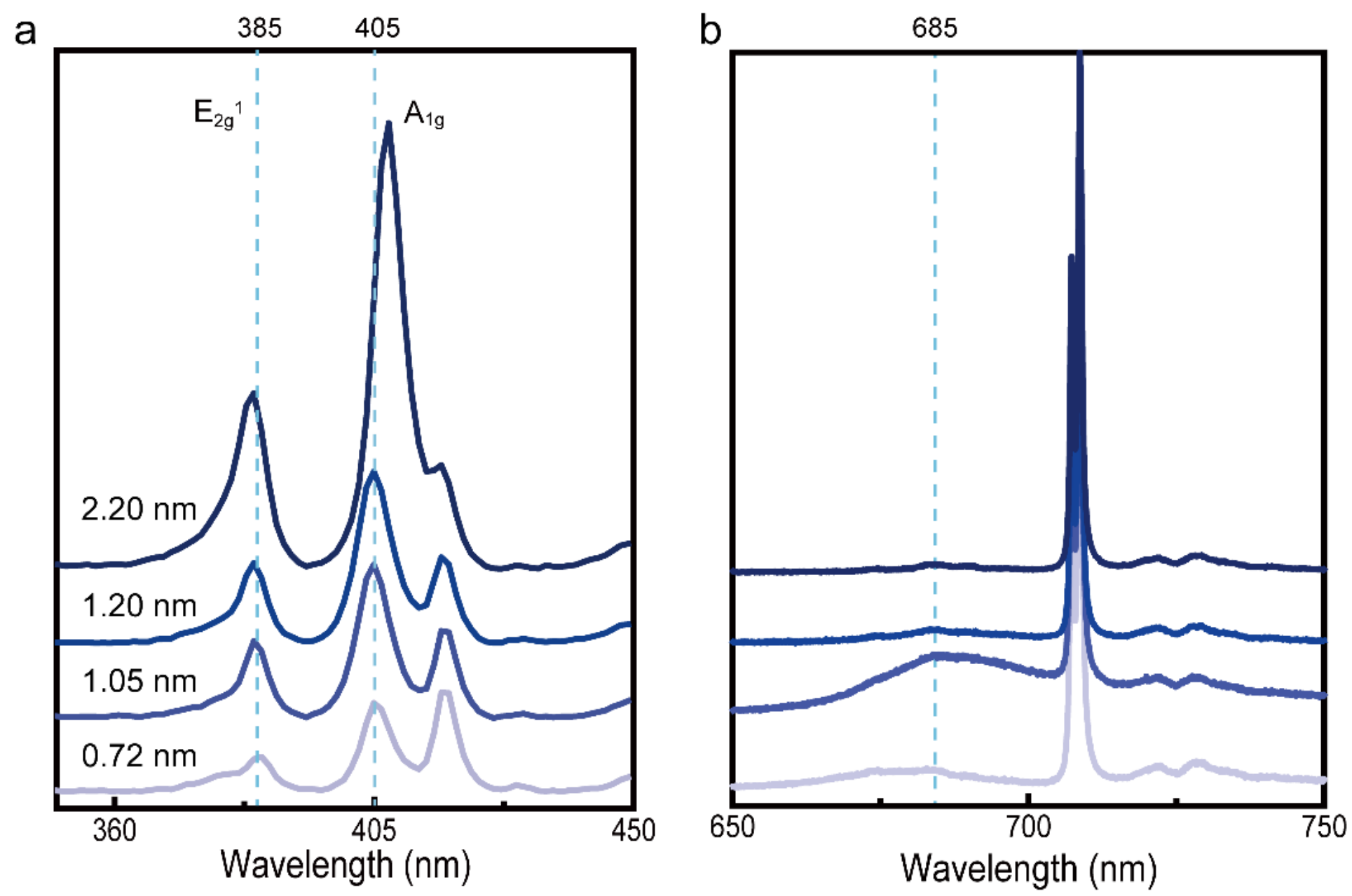

**Fig. S5. Raman and PL characterization of $MoS_2$ films with different thicknesses.** (a) Raman spectra of $MoS_2$ films with thicknesses of 0.72, 1.05, 1.20 and 2.20 nm, showing the characteristic $E_{2g}^{1}$ and $A_{1g}$ modes near 385 and 405$\text{cm}^{-1}$, respectively. (b) PL spectra of the corresponding $MoS_2$ films.

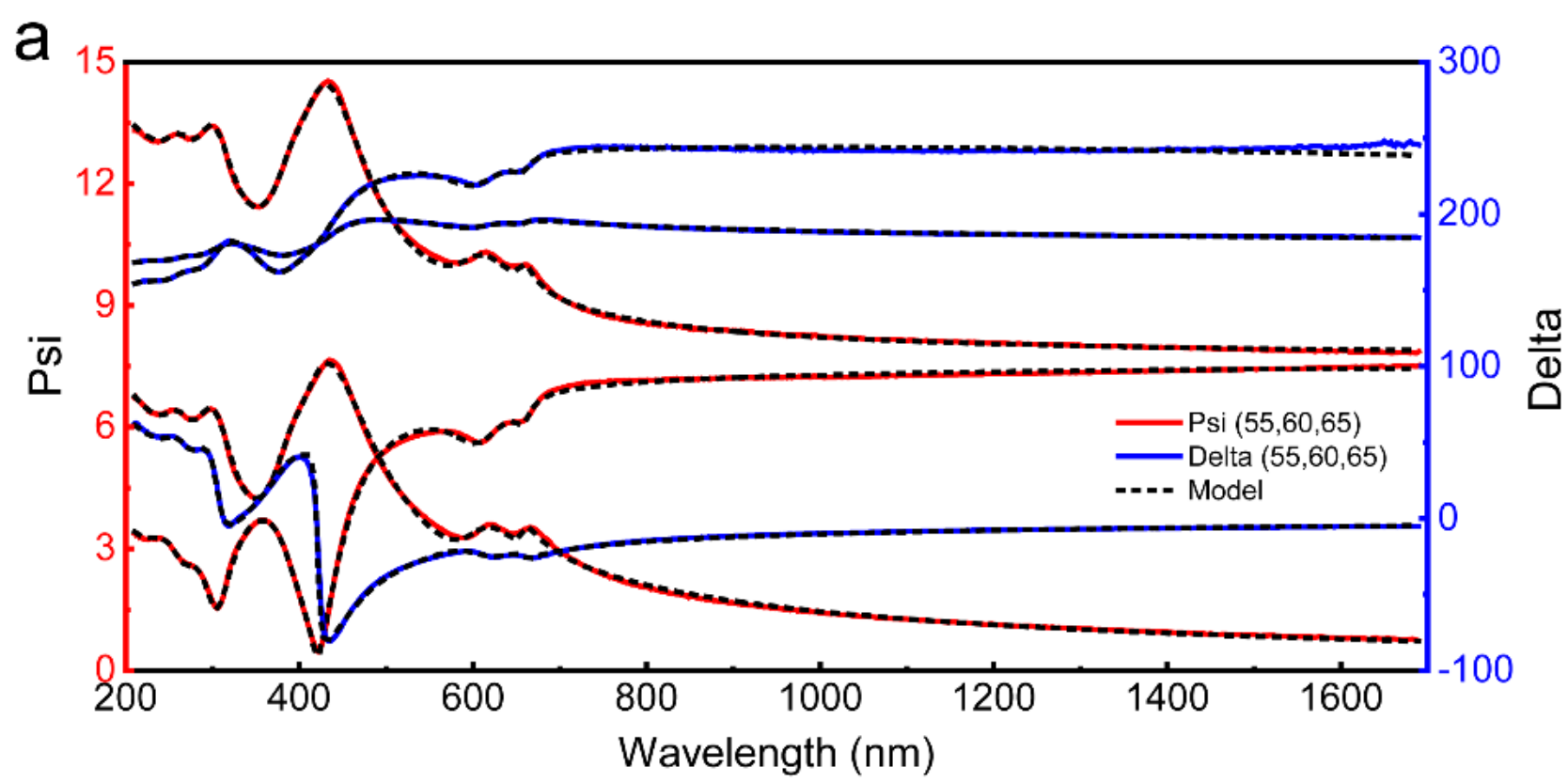

**Fig. S6. Extraction of optical constants of $MoS_2$ from SE measurements.** (a) Representative SE fitting results for the $MoS_2$ film. The experimental $\Psi$ and $\Delta$ spectra measured at incidence angles of 55°, 60° and 65° are well fitted by the optical model. Solid curves represent the measured spectra, and dashed black curves represent the fitted results.

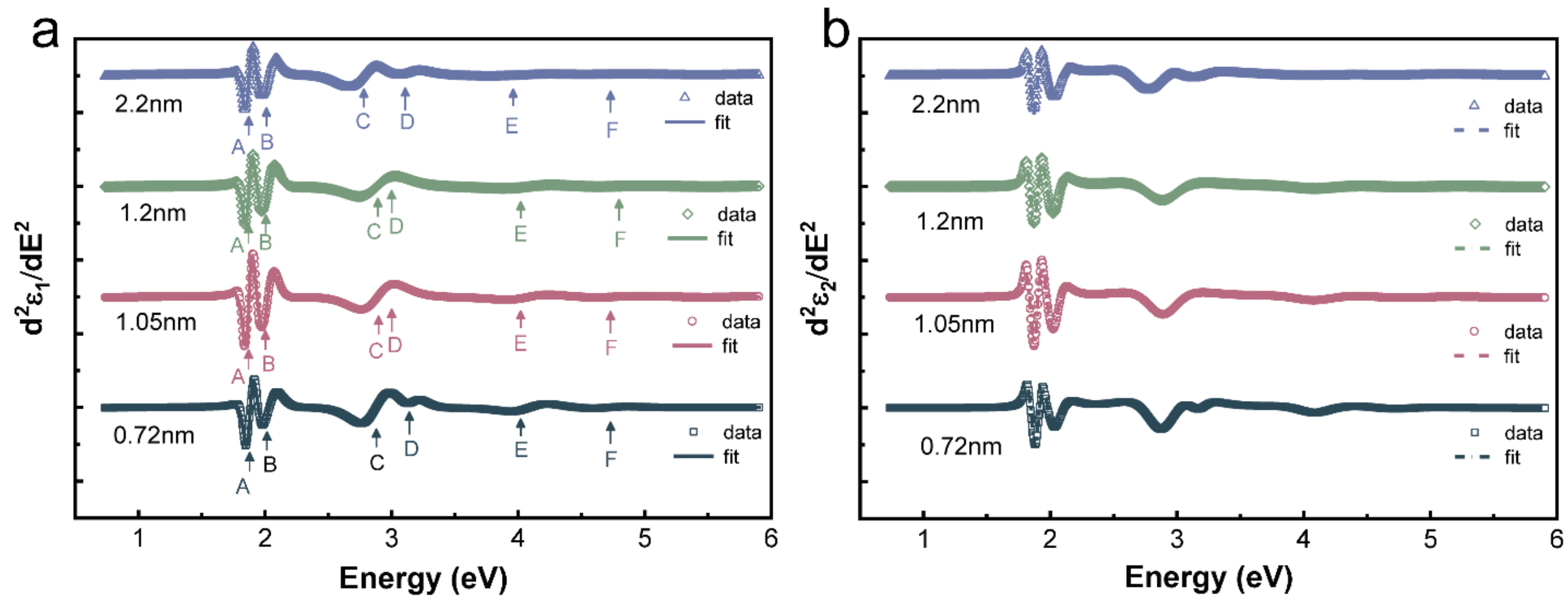

**Fig. S7. Critical point analysis.** Differently colored solid and dashed curves are the fits to the second-derivative spectra, plotted as symbols, of the real $\varepsilon_1$ and imaginary $\varepsilon_2$ parts of the dielectric function of $MoS_2$ film with different thickness.

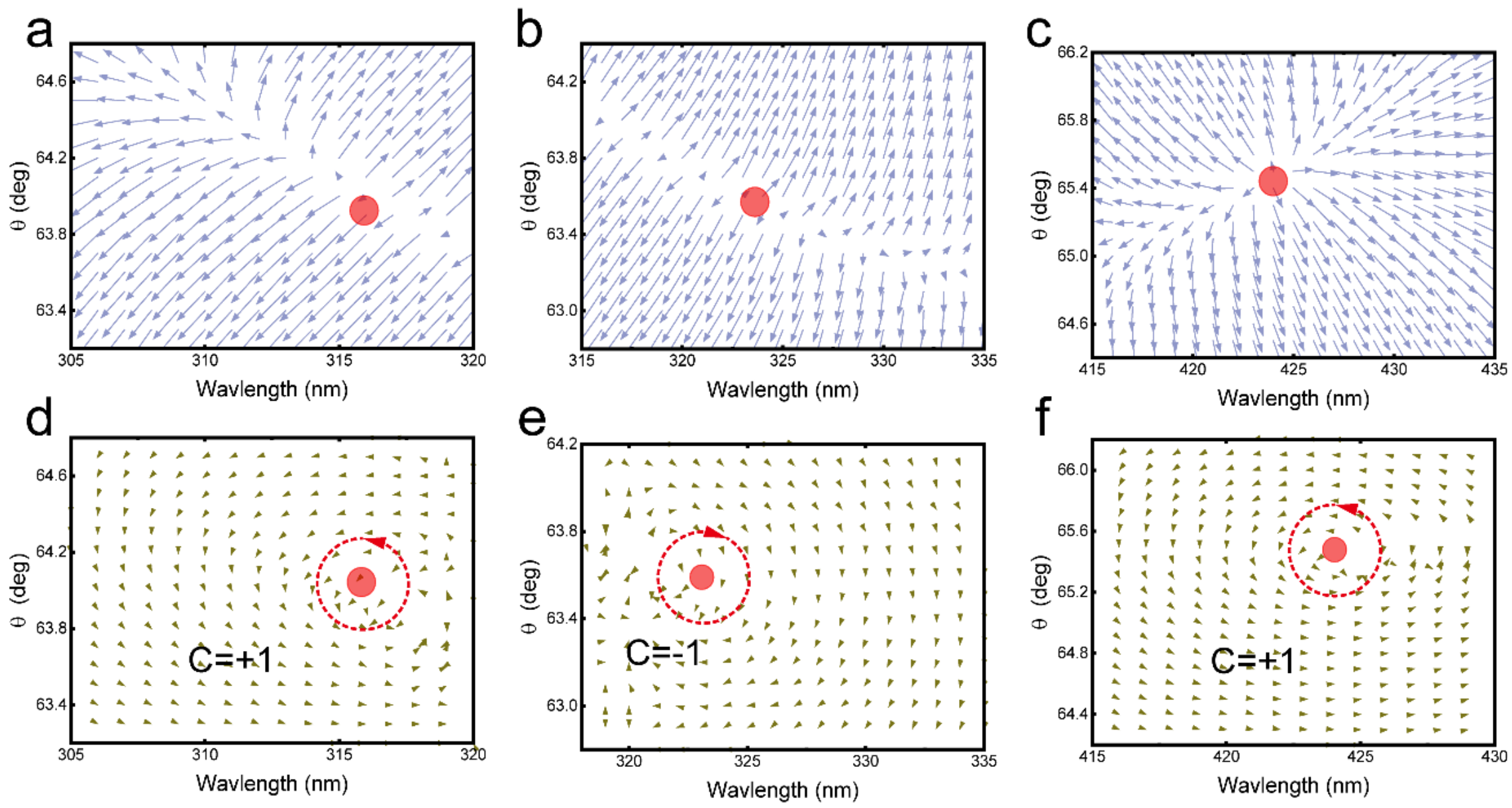

**Fig. S8. Gradient-field analysis of the measured Ψ and Δ maps of $MoS_2$ on sapphire.** (a–c) Gradient vector fields of measured Ψ maps highlight the three PSs marked by red dots. (d–f) Corresponding gradient vector fields of measured Δ maps. The circulating vector patterns around the singularities reveal the phase winding of the ellipsometric phase, corresponding to nontrivial topological charges of $C = +1$, $C = -1$, and $C = +1$, respectively.

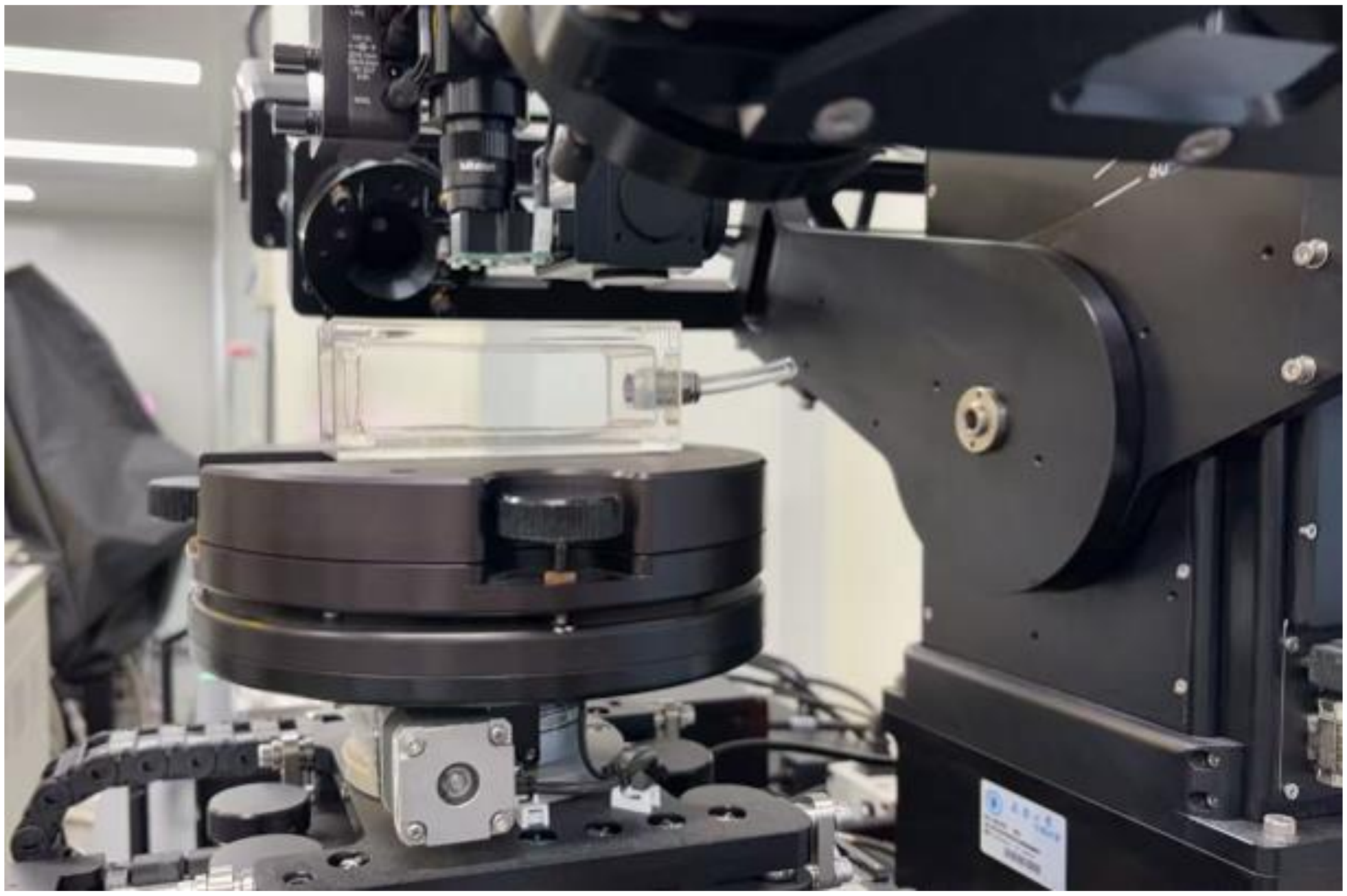

**Fig. S9. Sealed chamber for pressure-dependent refractive-index measurements.** Photograph showing the sealed chamber mounted on the ellipsometer stage. The $MoS_2$/sapphire sample was fixed inside the chamber with double-sided tape during the pressure-dependent SE measurements.

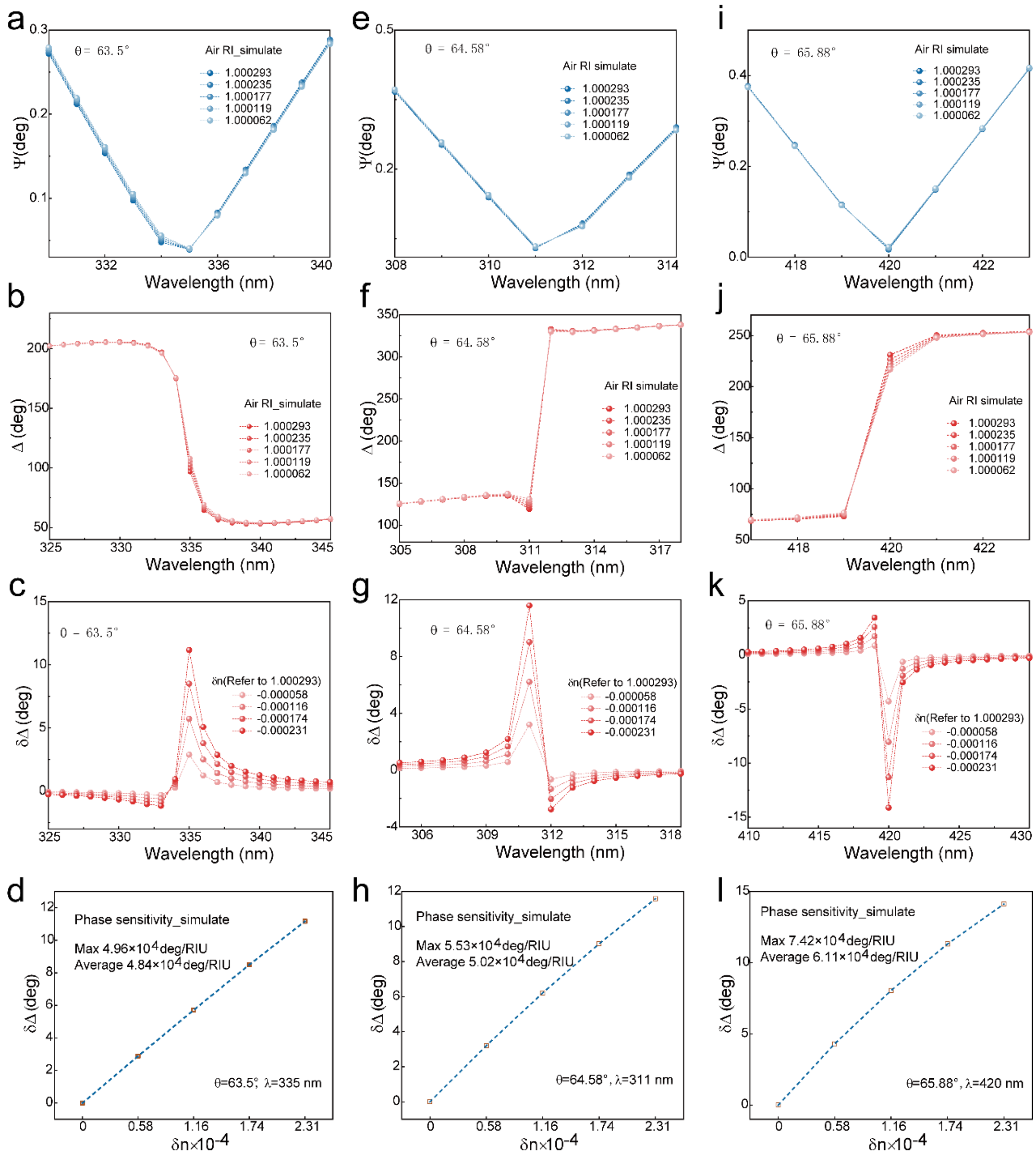

**Fig. S10. Simulated pressure-dependent air refractive-index response near the three PSs.** (a,e,i) Simulated $\Psi$ spectra near the three PSs for air refractive indices corresponding to gauge pressures from 0 to $-0.08$MPa. (b,f,j) Corresponding simulated $\Delta$ spectra. (c,g,k) Phase shifts $\delta\Delta$ referenced to 0 MPa gauge pressure. (d,h,l) Linear fits of $\delta\Delta$ versus $\delta n$ for extracting phase sensitivity. The three columns correspond to PSs near $\theta = 63.5°$, $64.58°$, and $65.88°$, respectively.

| Lorentz Oscillator | $A_j$ | $E_j$ (eV) | $\Gamma_j$ (eV) | $\varepsilon_\infty$ |
|---|---|---|---|---|
| 1(A) | 2.961631 | 1.882 | 0.0672 | 0.646 |
| 2(B) | 5.830530 | 2.036 | 0.2608 | |
| 3(C) | 24.333235 | 2.891 | 0.5600 | |
| 4(D) | 2.589626 | 3.171 | 0.3132 | |
| 5(E) | 8.253334 | 4.103 | 0.6880 | |
| 6(F) | 2.839092 | 4.724 | 0.7711 | |
| 7(G) | 6.025272 | 5.956 | 4.1066 | |

**Table S1. Fitting parameters of the dielectric functions of TMDCs monolayers.**

| CP energies (eV) | $0.72$ nm $MoS_2$ | $1.05$ nm $MoS_2$ | $1.2$ nm $MoS_2$ | $2.2$ nm $MoS_2$ |
|---|---|---|---|---|
| 1(A) | 1.8779 | 1.8695 | 1.8695 | 1.8699 |
| 2(B) | 2.014 | 2.0024 | 2.0085 | 2.0108 |
| 3(C) | 2.8796 | 2.8962 | 2.8908 | 2.7793 |
| 4(D) | 3.14 | 3 | 3 | 3.1077 |
| 5(E) | 4.02 | 4.02 | 4.02 | 3.96 |
| 6(F) | 4.73 | 4.73 | 4.7971 | 4.73 |

**Table S2. CP energies (eV) compared to different thickness MOS2 film.**

| CP broadening (eV) | 0.72 nm $MoS_2$ | 1.05 nm $MoS_2$ | 1.2 nm $MoS_2$ | 2.2 nm $MoS_2$ |
|---|---|---|---|---|
| 1(A) | 0.072114 | 0.06855 | 0.06882 | 0.068563 |
| 2(B) | 0.12744 | 0.10665 | 0.11061 | 0.12083 |
| 3(C) | 0.29226 | 0.31759 | 0.34682 | 0.28769 |
| 4(D) | 0.16466 | 1.1104 | 1.1786 | 0.27955 |
| 5(E) | 0.35612 | 0.39251 | 0.43926 | 0.67622 |
| 6(F) | 0.3189 | 0.39726 | 0.38768 | 0.44783 |

**Table S3. CP broadening (eV) compared to different thickness MOS2 film.**